\documentclass[prd,aps,tightenlines,floatfix,nofootinbib]{revtex4}
\usepackage{epsfig,graphicx}

\def\rmd{{\rm d}}
\def\etal{et al. }
\def\etals{et al.'s }
\def\msun{M$_{\odot}$}
\newcommand{\erf}[1]{(\ref{#1})}

\begin{document}

\title{Improved approximate inspirals of test-bodies into Kerr black holes}

\author{Jonathan R Gair}
\affiliation{Institute of Astronomy, Madingley Road, Cambridge, CB3 0HA, UK}

\author{Kostas Glampedakis} 
\affiliation{School of Mathematics, University of Southampton, Southampton SO17 1BJ, UK}

\date{\today}

%%%%%%%%%%%%%%%%%%%%%%%%%%%%%%%%%%%%%%%%%%%%%%%%%%%%%%%%%%%%%%%%%%%%%%%%%%%%%%%%%%%%%%%

\begin{abstract}

We present an improved version of the approximate scheme for generating inspirals of test-bodies into a Kerr black hole recently developed by Glampedakis, Hughes and Kennefick. Their original ``hybrid'' scheme was based on combining exact relativistic expressions for the evolution of the orbital
elements (the semi-latus rectum $p$ and eccentricity $e$) with approximate, weak-field, formula for the energy and angular momentum fluxes, amended by the assumption of constant inclination angle $\iota$ during the inspiral. Despite the fact that the resulting inspirals were overall well-behaved, certain pathologies remained for orbits in the strong-field regime and for orbits which are nearly circular and/or nearly polar. In this paper we eliminate these problems by incorporating an array of improvements in the approximate fluxes. Firstly, we add certain corrections which ensure the correct behaviour of the
fluxes in the limit of vanishing eccentricity and/or $90^\circ $ inclination. Secondly, we use higher order post-Newtonian formulae, adapted for generic orbits. Thirdly, we drop the assumption of constant inclination. Instead, we first evolve the Carter constant by means of an approximate post-Newtonian expression and subsequently extract the evolution of $\iota $. Finally, we improve the evolution of circular orbits by using fits to the angular momentum and inclination evolution determined by Teukolsky-based calculations. As an application of our improved scheme we provide a sample of generic Kerr inspirals which we expect to be the most accurate to date, and for the specific case of nearly circular orbits we locate the critical radius where orbits begin to decircularise under radiation reaction. These easy-to-generate inspirals should become a useful tool for exploring LISA data analysis issues and may ultimately play a role in the detection of inspiral signals in the LISA data.
\end{abstract}

\maketitle

%%%%%%%%%%%%%%%%%%%%%%%%%%%%%%%%%%%%%%%%%%%%%%%%%%%%%%%%%%%%%%%%%%%%%%%%%%%%%%%%%%%%%%

\section{Introduction}
\label{intro}

Among the most promising and rewarding sources of gravitational radiation for the future LISA space-observatory \cite{lisa} are the so-called extreme mass ratio inspirals (EMRIs). These binary systems are formed as the result of the capture of compact stellar remnants by supermassive black holes in the nuclei of galaxies \cite{capture}. As a result of two-body scattering in the surrounding cusp stellar population, such objects can end up on orbits that pass close to the central black hole. Occasionally these objects are captured by the massive black hole's gravitational ``pit'' with initial orbital parameters which ensure that the subsequent dynamics of the newly formed binary are almost entirely governed by gravitational radiation. The expected event rate for this scenario is quite promising (taking into account the capabilities of LISA), despite the uncertainties \cite{capture}, \cite{rates}, \cite{gair04}.   

The detection of EMRIs and the subsequent extraction for the system's parameters will rely heavily on the technique of matched filtering \cite{filter} which requires an accurate theoretical model of the true gravitational wave signal. The computation of EMRI waveforms is an ongoing effort, which is currently limited by our inability to accurately describe the orbital motion of the small body in the Kerr spacetime under the influence of radiation reaction. The traditional way of describing the system is via the celebrated Teukolsky perturbation formalism \cite{teuk} (where the system is modelled as a test-body in the field of a Kerr black hole) combined with the assumption of adiabatic evolution (see \cite{chapter},\cite{kg_review} for recent reviews and further references). This latter property is well justified as a first order approximation: the system's extreme mass ratio $\mu/M \sim 10^{-6} $ guarantees that any influence of gravitational back-reaction will become significant at timescales much longer than any orbital timescale. In the adiabatic approach one first assumes that the small body is moving on a geodesic and computes averaged fluxes of energy $E$ and angular momentum $L_z$ for this orbit. These fluxes are used to update the parameters of the geodesic and the procedure is then repeated. This procedure works very well in special cases: orbits that are either equatorial, or inclined and circular. For those two families the change 
in the third integral of motion, the Carter constant Q, is either trivial or can be directly inferred from the other two fluxes. This is no longer possible for generic (i.e., eccentric and inclined) orbits. Dealing with them 
requires input from the more advanced framework of gravitational self-force computations (see \cite{poisson}, \cite{cqg} for recent reviews in the field). 

Generating an inspiral and the associated waveform is a computationally expensive exercise even at the level of the Teukolsky formalism, let alone a self-force based calculation. Presently there are Teukolsky-based results for
equatorial or inclined-circular orbits that show how a given orbit would evolve under radiation reaction \cite{cutler}, \cite{scott}, \cite{kgdk} and recently some initial results have become available for generic orbits \cite{drasco05}. However, to date the only available full inspiral computation is the one by Hughes \cite{scott2} for circular-inclined orbits. A computation along the same lines can be carried out for equatorial-eccentric orbits and full generic inspirals should appear in the near future. However, these will be computationally very expensive.

Since present perturbative methods are either computationally demanding or incomplete it is highly desirable to have at hand a ``quick and dirty'' scheme for generating Kerr inspirals and waveforms, which can then be utilised in crucial data analysis computations for LISA \cite{gair04}. Glampedakis, Hughes and Kennefick \cite{GHK} (hereafter GHK) proposed a hybrid scheme for computing approximate generic EMRI trajectories. This hybrid scheme works by combining post-Newtonian radiation reaction formulae with a strong-field definition of the orbital parameters. It is able to reproduce many of the features expected from true inspirals and compares well to existing Teukolsky-based results \cite{cutler}, \cite{kgdk}, \cite{scott}. 

However, there are some significant problems with the approach in certain regions of the orbital phase space. We describe in the present paper a set of fixes and improvements to the GHK formalism that address these problems. The most crucial improvement comes from the study of the limiting cases of nearly circular and/or nearly polar orbits. In order to guarantee a smooth orbital evolution, the exact fluxes need to satisfy certain consistency relations. By adding appropriate correction terms we can enforce the fluxes to satisfy these relations, thereby eliminating the pathologies of the GHK scheme in those regions of phase space. 

We also make improvements to the weak-flux expressions for $E$ and $L_z$. An immediate improvement comes from simply using available higher post-Newtonian (PN) expressions. By combining Tagoshi's \cite{tagoshi} 2.5PN fluxes for equatorial small-eccentricity orbits, Shibata {\it et al.}'s \cite{shibata} 2PN fluxes for circular small-inclination orbits and Ryan's \cite{Ryan} fluxes (which are fully accurate in the leading Newtonian and 1.5PN pieces ), we construct approximate 2PN fluxes which perform well for practically all eccentricities and inclinations and improve the strong-field behaviour of the GHK inspirals.  

A further modification is to change the prescription for the evolution of the orbital inclination, or equivalently for the Carter constant flux. GHK adopted the simple (but fairly accurate) rule of fixed inclination during inspiral. Here we improve on this by adding the next order spin-dependent correction, and allowing inclination to evolve. A final improvement to the fluxes comes by fitting functions to the results of Teukolsky-based computations for circular-inclined orbits \cite{scott}. The combination of the above ingredients ensures the new hybrid scheme is applicable throughout parameter space, is qualitatively correct everywhere, and exhibits very good performance when compared to accurate, Teukolsky-based inspirals.

The paper is organised as follows. In section~\ref{GHKScheme} we give a brief overview of the GHK approach to computing inspirals. In sections~\ref{nearcirc}--\ref{highinc} we describe two corrections that are required to give physically reasonable behaviour in nearly circular and nearly polar inspirals. These corrections must be included when any approximation is 
used for the energy and angular momentum fluxes. Then, in sections~\ref{2PN}--\ref{newQdot}, we provide new flux approximations that are accurate to higher post-Newtonian order than the leading order fluxes used in \cite{GHK}. In section~\ref{applications} we illustrate the use of these expressions by investigating the properties of some example inspirals. Finally, in section~\ref{conscorr}, we briefly discuss the conservative component of the self-force which is not otherwise considered in this analysis. In the standard manner we adopt geometrised units $G=c=1$.

%%%%%%%%%%%%%%%%%%%%%%%%%%%%%%%%%%%%%%%%%%%%%%%%%%%%%%%%%%%%%%%%%%%%%%%%%%%%%%%%%%%%%%%%%%%%%%%%%%

\section{Summary of GHK hybrid scheme}
\label{GHKScheme}
In the GHK paper \cite{GHK}, the orbit is reparameterized in terms of an eccentricity, $e$, semi-latus rectum, $p$, and 
inclination angle, $\iota$. The parameters $e$ and $p$ are defined in terms of the turning points of the radial motion, which 
are determined by the roots of the radial potential
\begin{equation}
R(r) \equiv \left[E\,(r^{2}+a^{2})-a\,L_{z}\right]^{2}-\left(r^{2}-2\,M\,r+a^{2}\right)\,\left[\mu^{2}\,r^{2}+\left(L_{z}
-a\,E\right)^{2}+Q\right].
\label{radpot}
\end{equation}
We use $M$ and $\mu$ to denote the masses of the primary and secondary respectively. For a given value of the energy, $E$, $z$-component of the angular momentum, $L_{z}$, and Carter constant, $Q$, 
the motion oscillates between a periapse, $r_p$, and an apoapse, $r_{a}$. Once these have been determined from \erf{radpot}, 
$p$ and $e$ are defined by
\begin{equation}
r_{p}=\frac{p}{1+e}, \qquad r_{a}=\frac{p}{1-e} \qquad \Rightarrow \qquad p = \frac{2\,r_{a}\,r_{p}}{r_{a}+r_{p}}, 
\qquad e=\frac{r_{a}-r_{p}}{r_{a}+r_{p}}.
\label{pedef}
\end{equation}
The orbital inclination is defined in terms of the Carter constant by
\begin{equation}
Q=L_{z}^{2}\,\tan^{2}\iota.
\label{idef}
\end{equation}
In the limit $a\rightarrow 0$, $Q \rightarrow L_{x}^{2}+L_{y}^{2} = L^{2}-L_{z}^{2}$, and definition \erf{idef} agrees with 
the usual notion of inclination.

GHK constructed inspirals by evolving $E$, $L_{z}$ and $Q$. The energy and angular momentum were evolved using lowest 
order post-Newtonian fluxes, modified from Ryan \cite{Ryan}
\begin{eqnarray}
\dot{E}_R &=& -\frac{32}{5}\frac{\mu^2}{M^2}
\left ( \frac{M}{p} \right )^{5} (1-e^2)^{3/2}
\left [ f_{1}(e)  - \frac{a}{M} \left ( \frac{M}{p} \right )^{3/2}
\cos\iota f_{2}(e) \right ]\;,
\label{GHKEdot} \\
\dot{L}_{z\,R} &=& -\frac{32}{5}\frac{\mu^2}{M}
\left (\frac{M}{p} \right )^{7/2}
(1-e^2)^{3/2} \left[ \cos\iota f_{3}(e) + \frac{a}{M}
\left ( \frac{M}{p} \right )^{3/2}
\left[f_{4}(e) -\cos^2 \iota f_{5}(e)\right] \right]\;,
\label{GHKLdot}
\end{eqnarray}
in which
\begin{eqnarray}
f_{1}(e) &=& 1+ \frac{73}{24}e^2 + \frac{37}{96} e^4\;,
\qquad
f_{2}(e) =  \frac{73}{12} + \frac{823}{24}e^2 +
\frac{949}{32} e^4 + \frac{491}{192} e^6\;,
\qquad
f_{3}(e) =  1+ \frac{7}{8}e^2\;, \nonumber
\\
f_{4}(e) &=& \frac{61}{24} + \frac{63}{8} e^2 + \frac{95}{64} e^4\;,
\qquad
f_{5}(e) = \frac{61}{8} + \frac{91}{4}e^2 + \frac{461}{64} e^4\;,
\label{GHKefactors}
\end{eqnarray}
Ryan also provides an expression for the evolution of $Q$, but this was found to overpredict the change in inclination angle, 
when compared to accurate Teukolsky-based calculations for circular-inclined orbits \cite{scott}. For this reason, GHK used 
a ``constant inclination'' approximation\footnote{This rule is in fact exact in a spherically symmetric gravitational field,
see \cite{GHK} for a discussion.} to evolve $Q$
\begin{equation}
\dot{\iota} = 0 \qquad \Rightarrow \qquad \dot{Q}_{\rm sph} = \frac{2\,Q}{L_{z}}\,\dot{L}_{z} \;.
\label{GHKidot}
\end{equation}
The energy and angular momentum fluxes (\ref{GHKEdot}--\ref{GHKLdot}) are evaluated using the $p$ and $e$ given by definition 
\erf{pedef}. This was found to work much better in the strong-field than re-writing the right hand sides of 
\erf{GHKEdot}--\erf{GHKLdot} using the Keplerian relation between $E$, $L_{z}$ and $p$, $e$ and evaluating the resulting 
expressions for the orbit with the same $E$, $L_{z}$ and $Q$.

%%%%%%%%%%%%%%%%%%%%%%%%%%%%%%%%%%%%%%%%%%%%%%%%%%%%%%%%%%%%%%%%%%%%%%%%%%%%%%%%%%%%%%%%%%%%%%%%%%%%%%%%%%%%%%%%%%%%%%%%%%%%

\section{Correcting the behaviour of nearly circular inspirals}
\label{nearcirc}
The original GHK inspirals exhibit some unphysical features for nearly circular ($e \approx 0$) and nearly polar 
($\iota \approx \pi/2$) inspirals. These problems arise because the GHK approach uses post-Newtonian expressions to evolve 
$E$, $L_{z}$ and $Q$, which are only valid to certain orders in the spin, eccentricity etc., but then computes the evolution 
of the eccentricity and semi-latus rectum using the {\it exact} transformation law, accurate to arbitrary order. As a consequence, 
sensitive cancellations which are needed to ensure reasonable behaviour in certain limits do not occur and this leads to some 
bizarre behaviour. This can be ameliorated somewhat by extending the flux expressions to higher post-Newtonian order 
(see section~\ref{2PN}), but the problems are not eliminated entirely. However, it is possible to make corrections to the fluxes 
that force the necessary cancellations to occur. We describe these corrections in this and the following section.

\begin{figure}
\centerline{\includegraphics[keepaspectratio=true,height=5in,angle=-90]{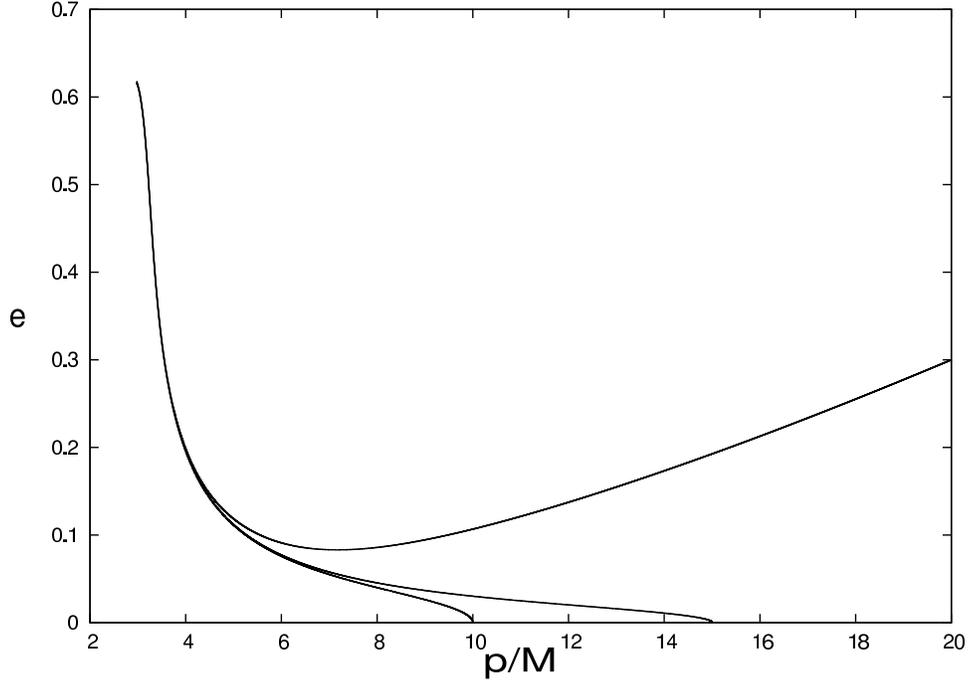}}
\caption{Illustration of the problem with the hybrid fluxes for nearly circular orbits. We show three inspirals in the equatorial plane of a black hole with spin $a= 0.9 M$.}
\label{circprob}
\end{figure}

In Figure~\ref{circprob} we show a selection of inspirals in the equatorial plane of a black hole with spin parameter 
$a=0.9 {\rm M}$. Three inspirals are shown, with initial semi-latus rectums and eccentricities given by the pairs 
$(20\,{\rm M}, 0.3)$, $(15\,{\rm M},0.001)$ and $(10\,{\rm M},0.001)$. The inspirals have been computed using the original GHK 
fluxes \erf{GHKEdot}--\erf{GHKLdot}. The unphysical behaviour near $e=0$ manifests itself in two ways. For inspirals that start 
with comparatively large semi-latus rectum, but very small eccentricity, the inspiral moves rapidly away from circularity -- 
in fact, the trajectories are almost vertical near to the $e=0$ axis. For inspirals that start with high eccentricity, the 
turnover point at which the eccentricity starts to increase occurs at a higher value of $p/M$ than one would expect from 
Teukolsky-based calculations \cite{kenn98}. There is also a strange tendency for the curves to attract -- for these quite 
diverse choices of initial parameters, every inspiral plunges at nearly the same point in phase space. Changing from the 
original GHK fluxes to the higher order fluxes described in section~\ref{2PN} alleviates the rapid increase in eccentricity 
prior to plunge, but the nearly circular orbits are still pushed away from circularity and the trajectories still converge at 
plunge.

%%%%%%%%%%%%%%%%%%%%%%%%%%%%%%%%%%%%%%%%%%%%%%%%%%%%%%%%%%%%%%%%

\subsection{The cause of the problem}
The origin of this problem can be understood by considering the behaviour of the fluxes near $e=0$. The rate of change of 
the eccentricity is given by the fluxes of energy, angular momentum and the rate of change of the inclination angle, 
$\dot{\iota}$ (or equivalently, the flux of the Carter constant $\dot{Q}$) by
\begin{equation}
\dot{e}  = \frac{\partial e}{\partial E} \, \dot{E} + \frac{\partial e}{\partial L_{z}} \, \dot{L}_{z} + 
\frac{\partial e}{\partial \iota} \, \dot{\iota} .
\label{edot}
\end{equation}
The partial derivatives may be derived from the definitions of the radial potential \erf{radpot} and eccentricity 
\erf{pedef}
\begin{eqnarray}
\nonumber R(r) &=& (E^2 -\mu^2)\,(r-r_a)(r- r_p)(r-r_3)(r-r_4)
%\left[E\,(r^{2}+a^{2})-a\,L_{z}\right]^{2}&-&\left(r^{2}-2\,M\,r+a^{2}\right)\,\left[\mu^{2}\,r^{2}+
%\left(L_{z}-a\,E\right)^{2}+L_{z}^{2}\, \tan^{2}\iota\right] 
\\ && =(\mu^{2}-E^{2})\,\left(-r^{2}+
\frac{2\,p}{1-e^{2}}\,r-\frac{p^{2}}{1-e^{2}}\right)\,(r-r_{3})\,(r-r_{4})
\label{edef}
\end{eqnarray}
Since the eccentricity enters this equation only as $e^{2}$, it follows that $E$, $L_{z}$ and the fluxes $\dot{E}$ 
and $\dot{L}_{z}$ must be functions of $p$, $e^{2}$ and $\iota$. The post-Newtonian fluxes \erf{GHKEdot}--\erf{GHKLdot} 
satisfy this condition. However, it also means that $e^{2}$ may be written as a function of $E$ and $L_{z}$. We anticipate 
that $\partial(e^{2})/\partial E$ is well behaved for all eccentricities, but 
$\partial e/\partial E = (\partial(e^{2})/\partial E)/e$ will diverge like $1/e$. Differentiation of the condition 
\erf{edef} yields the necessary partial derivatives
\begin{eqnarray}
e\,\frac{\partial e}{\partial E} &=& \frac{e\,(1-e^{2})}{2\,p}\,\left((1+e)\,\frac{N_{1}\left(r_p\right)}{D
\left(r_p \right)} - (1-e)\,\frac{N_{1}\left(r_a\right)}{D\left(r_a\right)} \right), 
\nonumber \\
\nonumber \\
e\,\frac{\partial e}{\partial L_{z}} &=& \frac{e\,(1-e^{2})}{2\,p}\,\left((1+e)\,\frac{N_{2}\left(r_p\right)}{D
\left(r_p\right)} - (1-e)\,\frac{N_{2}\left(r_a \right)}{D\left(r_a \right)} \right), 
\nonumber \\
\nonumber \\
e\,\frac{\partial e}{\partial \iota} &=& \frac{e\,(1-e^{2})}{2\,p}\,\left((1+e)\,\frac{N_{3}\left(r_p \right)}
{D\left(r_p \right)} - (1-e)\,\frac{N_{3}\left(r_a \right)}{D\left(r_a \right)} \right), 
\nonumber \\
\nonumber \\
{\rm where} \qquad N_{1}(r) &=& E\,r^{4}+a^{2}\,E\,r^{2}-2\,a\,M\,(L_{z}-a\,E)\,r, 
\nonumber \\
\nonumber \\ 
N_{2}(r) &=& -L_{z}\,\sec^{2}\iota\,r^{2}+2\,M\,(L_{z}\,\sec^{2}\iota-a\,E)\,r -a^{2}\,L_{z}\,\tan^{2}\iota 
\nonumber \\
\nonumber \\ 
N_{3}(r) &=& L_{z}^{2}\,\tan\iota\,\sec^{2}\iota\,(r\,(2\,M-r)-a^{2}), 
\nonumber \\
\nonumber \\ 
D(r) &=& -2\,(\mu^{2}-E^{2})\,r^{3}+3\,\mu^{2}\,M\,r^{2}-(a^{2}\,(\mu^{2}-E^{2})+L_{z}^{2}\,\sec^{2}\iota)\,r 
\nonumber \\
&& + L_{z}^{2} \tan^{2}\iota\,M + (L_{z}-a\,E)^{2}\,M 
\label{ederivs}
\end{eqnarray}
The substitution $e \leftrightarrow -e$ leaves the right hand side of these expressions unchanged, so as expected 
the derivatives are functions of $e^{2}$ only. The function $D(r)$ is given by the derivative of the radial potential 
$\frac{1}{2}\,\partial R/\partial r$ and so we can re-write it
\begin{eqnarray}
D(r)&=&-\frac{1}{2} \, (\mu^{2}-E^{2}) \left[-\left(r -r_p \right)\,(r-r_{3})\,(r-r_{4}) +\left(r_a-r\right)\,
(r-r_{3})\,(r-r_{4}) \nonumber \right. \\ && \left. + \left(r_a-r\right)\,\left(r-r_p \right)\,(r-r_{4}) 
+ \left(r_a  -r\right)\,\left(r -r_p \right)\,(r-r_{3})\right] \nonumber \\
\Rightarrow D\left( r_p \right) &=& \frac{e\,p}{1-e^{2}}\,(\mu^{2}-E^{2}(p,\iota,e))\,\left(r_p
-r_{3}(p,\iota,e)\right)\,\left(r_p-r_{4}(p,\iota,e)\right) \nonumber \\
D\left( r_a \right) &=& -\frac{e\,p}{1-e^{2}}\,(\mu^{2}-E^{2}(p,\iota,e))\,
\left(r_a  -r_{3}(p,\iota,e)\right)\,\left(r_a  -r_{4}(p,\iota,e)\right)
\end{eqnarray}
In the limit of small eccentricity at fixed $p$ and $\iota$, we deduce
\begin{eqnarray}
\lim_{e\rightarrow0} \left(e\,\frac{\partial e}{\partial E} \right) &=& \frac{N_{1}(p,\iota)}{p^{2}\,
(\mu^{2}-E^{2}(p,\iota,0))\,(p-r_{3}(p,\iota,0))\,(p-r_{4}(p,\iota,0))} + {\cal O}(e^{2}), 
\nonumber \\ 
\lim_{e\rightarrow0} \left(e\,\frac{\partial e}{\partial L_{z}} \right) &=& \frac{N_{2}(p,\iota)}{p^{2}\,
(\mu^{2}-E^{2}(p,\iota,0))\,(p-r_{3}(p,\iota,0))\,(p-r_{4}(p,\iota,0))} + {\cal O}(e^{2}), \nonumber \\
\lim_{e\rightarrow0} \left(e\,\frac{\partial e}{\partial \iota} \right) &=& \frac{N_{3}(p,\iota)}{p^{2}\,
(\mu^{2}-E^{2}(p,\iota,0))\,(p-r_{3}(p,\iota,0))\,(p-r_{4}(p,\iota,0))} + {\cal O}(e^{2})
\end{eqnarray}
where $N_1(p,\iota)=N_1(r_a)=N_2(r_p) $ for $e=0$ and similarly for $N_2,N_3 $. 

It is now clear from \erf{edot} that $\dot{e} \propto 1/e$ for small $e$, unless the fluxes satisfy a consistency 
condition in the limit $e\rightarrow 0$
\begin{equation}
N_{1}(p,\iota)\,\dot{E} \,(p,\iota,0) + N_{2}(p,\iota)\,\dot{L}_{z} \,(p,\iota,0) + 
N_{3}(p,\iota)\,\dot{\iota} \,(p,\iota,0) = 0 \label{circcond}
\end{equation}
We will see in section~\ref{highinc} that it can be better to evolve the Carter constant, $Q$, instead of $\iota$. Using the definition \erf{idef} to write $\dot{\iota}$ in terms of $\dot{Q}$ and $\dot{L_z}$, \erf{circcond} 
becomes
\begin{equation}
N_{1}(p,\iota)\,\dot{E} \,(p,\iota,0) + N_{4}(p,\iota)\,\dot{L}_{z} \,(p,\iota,0) + N_{5}(p,\iota)\,\dot{Q} \,(p,\iota,0) = 0 
\label{circcondQform}
\end{equation}
where
\begin{equation}
N_4(p,\iota) = (2\,M\,p-p^2)\,L_z - 2\,M\,a\,E\,p, \qquad N_5 (p,\iota) = (2\,M\,p-p^2-a^2)/2 . \label{QNdefs}
\end{equation}
It is understood that the $N_i$'s are evaluated for the circular orbit with the specified $p$ and $\iota$. 
The GHK fluxes \cite{GHK} do {\it not} satisfy this relation, which explains the bizarre behaviour seen for small eccentricity. 

What does the condition \erf{circcond} or \erf{circcondQform} mean physically? For simplicity, we first examine equatorial 
orbits before extending to inclined orbits. In the equatorial plane, $\dot{\iota}=0$ by symmetry, and this condition is 
satisfied by the GHK scheme \erf{GHKidot}. In the circular equatorial limit, the energy and angular momentum are given in 
terms of the semi-latus rectum by
\begin{equation}
\frac{E}{\mu}=\frac{1-2\,(M/p)\pm (a/M)\,\left(M/p\right)^{\frac{3}{2}}}{\sqrt{1-3\,(M/p) \pm 2\,
(a/M)\,\left(M/p\right)^{\frac{3}{2}}}}, \qquad \frac{L_{z}}{\mu\,M} = 
\pm \, \left(\frac{p}{M}\right)^{\frac{1}{2}} \, \frac{1\mp 2\,(a/M)\,
\left(M/p\right)^{\frac{3}{2}}+\left(a/M\right)^{2}\, \left(M/p\right)^{2}}{\sqrt{1-3\,(M/p) 
\pm 2\,(a/M)\,\left(M/p\right)^{\frac{3}{2}}}}
\end{equation}
Where the upper (lower) sign is for prograde (retrograde) orbits. We find that condition \erf{circcond} reduces to
\begin{equation}
\dot{E} (p,0/\pi,0) = \pm\, \frac{\sqrt{M}}{p^{\frac{3}{2}}\pm a\,\sqrt{M}} \, \dot{L}_{z} (p,0/\pi,0) = 
\Omega_{\phi} (p)\, \dot{L}_{z} (p,0/\pi,0)
\label{eqcond}
\end{equation}
where $\Omega_{\phi} (p)$ is the azimuthal velocity, $\rmd \phi/\rmd t$, of the circular orbit. Expression \erf{eqcond} 
is just the condition that circular orbits remain circular under radiation reaction, which must be true for the 
actual gravitational waves emitted by the orbit \cite{dan}. The general condition \erf{circcond} is similarly 
just another way to write the circular goes to circular rule \cite{dan}. By forcing the leading term to cancel 
from our $\dot{e}$ expression, we ensure that $\dot{e} \propto e$ and therefore that $\dot{e}=0$ for circular orbits. 

We can compute the difference $\dot{E} - \Omega_{\phi} \, \dot{L_{z}}$ for circular equatorial orbits under 
the GHK scheme. Expanding in powers of the spin parameter, $a$, we find
\begin{eqnarray}
\dot{E} - \Omega_{\phi} \, \dot{L}_{z} &=& -\frac{32}{5}\,\frac{\mu^{2}}{M^{2}}\,\left(\frac{M}{p}\right)^{5}\,
\left(1\mp\frac{73}{12}\,\frac{a\,\sqrt{M}}{p^{\frac{3}{2}}}\right)+\frac{M^{\frac{3}{2}}}{p^{\frac{3}{2}}+a\,
\sqrt{M}}\,\frac{32}{5}\,\frac{\mu^{2}}{M^{2}}\,\left(\frac{M}{p}\right)^{\frac{7}{2}}\, \left(1\mp\frac{61}{12}\,
\frac{a\,\sqrt{M}}{p^{\frac{3}{2}}}\right) \nonumber \\ &=& {\cal O}\left( \frac{a^2\,
M}{p^{3}}\right)
\end{eqnarray}
The cancellation occurs correctly to linear order in the spin. This is expected since the fluxes were derived in a low-spin limit. If we use the higher order fluxes \erf{new_Edot}--\erf{new_Ldot} described in section~\ref{2PN}, we similarly find
\begin{eqnarray}
\dot{E} - \Omega_{\phi} \, \dot{L}_{z} &=& -\frac{32}{5}\,\frac{\mu^{2}}{M^{2}}\,\left(\frac{M}{p}\right)^{5}\,
\left(1\mp\frac{73}{12}\,\frac{a\,\sqrt{M}}{p^{\frac{3}{2}}}-\frac{1247}{336}\,\frac{M}{p}+4\,\pi\,
\left(\frac{M}{p}\right)^{\frac{3}{2}} -\frac{44711}{9072} \, \left(\frac{M}{p}\right)^{2} +\frac{33}{16}\,
\left(\frac{a}{p}\right)^{2}\right) \nonumber \\ && \hspace{0.05in} +\frac{1}{p^{\frac{3}{2}}+a\,\sqrt{M}}\,
\frac{32}{5}\,\frac{\mu^{2}\,M^{3}}{p^{\frac{7}{2}}}\, \left(1\mp\frac{61}{12}\,\frac{a\,
\sqrt{M}}{p^{\frac{3}{2}}} -\frac{1247}{336}\,\frac{M}{p}+4\,\pi\,\left(\frac{M}{p}\right)^{\frac{3}{2}} -
\frac{44711}{9072} \,\left(\frac{M}{p}\right)^{2} +\frac{33}{16}\,\left(\frac{a}{p}\right)^{2} \right) 
\nonumber \\ &=& {\cal O} \left(\frac{a^2\,M^{\frac{3}{2}}}{p^{\frac{5}{2}}}\right)
\end{eqnarray}
In each case, the cancellation occurs to the post-Newtonian order at which the fluxes were derived. However, when we use the fluxes to evolve inspirals, this is no longer accurate enough since we are using definitions of eccentricity and semi-latus rectum that are accurate to arbitrary order. Since we know that the circular goes to circular rule must hold for real inspirals, we can improve our approximate scheme by modifying our fluxes to enforce this condition.

%%%%%%%%%%%%%%%%%%%%%%%%%%%%%%%%%%%%%%%%%%%%%%%%%%%%%%%%%%%%%%%%%%%%%%%%%%%%%%%%%%%%%%%%%%%%%%%%%%%%%%%%%%%%%%%%%%%%%%%%%%%%%

\subsection{Correcting the hybrid fluxes}
The condition \erf{circcond} is a relationship between the fluxes for circular orbits. The easiest way to impose it is to 
specify two of the flux expressions and use \erf{circcond} to determine the other. Since this relation applies only to circular 
orbits, we adopt the approach of leaving the eccentricity corrections largely unchanged, and correct only the circular piece 
of the fluxes. However, we must include the usual $(1-e^{2})^{\frac{3}{2}}$ factor in front of the near-circular correction 
to ensure reasonable behaviour at high eccentricity. As the eccentricity approaches the parabolic limit, $e=1$, at fixed 
semi-latus rectum, we expect the total change in $E$, $L_{z}$, $Q$ and $\iota$ to be approximately constant, since the 
strong-field part of the orbit, where most of the radiation is lost, does not change significantly. However, the orbital 
period diverges like $(1-e^{2})^{-\frac{3}{2}}$, and therefore the orbital averaged flux must tend to zero in the same way. 
This is the origin of the $(1-e^{2})^{\frac{3}{2}}$ prefactor in \erf{GHKEdot}--\erf{GHKLdot}, and we must also include it 
in the circular correction. We cannot reliably say anything about further eccentricity corrections, since such corrections 
are higher order than the available post-Newtonian results. Our approach ensures that the eccentricity pieces are 
asymptotically correct and the new fluxes have the right leading order behaviour at high eccentricity. It may be possible 
to get more accurate expressions by applying the correction to the other eccentricity terms as well, but a priori we have 
no justification for doing this.

\begin{figure}
\centerline{\includegraphics[keepaspectratio=true,height=5in,angle=-90]{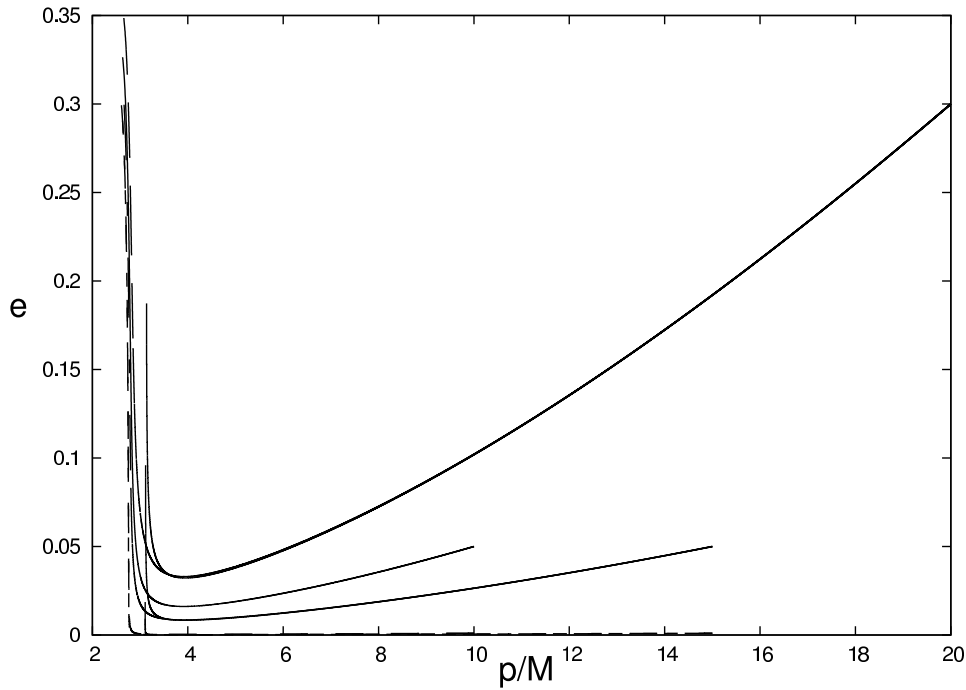}}
\caption{Improved inspiral trajectories after implementation of the fix described in the text. We show inspirals in 
the equatorial plane of a black hole with spin $a=0.9M$. The solid lines correspond to using the post-Newtonian 
$\dot{E}$ expression \erf{GHKEdot} to correct the angular momentum flux, $\dot{L}_{z}$. The dashed lines 
correspond to using the post-Newtonian $\dot{L}_{z}$ expression \erf{GHKLdot} to correct the energy flux, $\dot{E}$.}
\label{circfix}
\end{figure}

If we chose to fix $\dot{E}$ and $\dot{L}_z$ by expressions \erf{GHKEdot}--\erf{GHKLdot} and use \erf{circcond} to 
determine $\dot{\iota}$, we would find that $\dot{\iota} \neq 0$ in the equatorial plane, which is unphysical. 
For simplicity, we want to use the same correction throughout parameter space, so we will modify either 
$\dot{E}$ or $\dot{L}_{z}$. In Figure~\ref{circfix} we show new equatorial inspirals computed by imposing 
the fix in two different ways. For the solid lines we use the post-Newtonian $\dot{E}$ expression \erf{GHKEdot}, 
and correct the circular piece of the post-Newtonian $\dot{L}_{z}$ using \erf{circcond}. For the dashed lines, 
we leave the post-Newtonian $\dot{L}_z$ expression \erf{GHKLdot} unchanged, and correct the circular piece 
of the $\dot{E}$ expression. We note first that it does not make a great deal of difference which way the 
fix is implemented. The trajectories differ a little near to plunge, but are largely the same. The most striking 
feature of Figure~\ref{circfix} is how much better the inspirals now look. The orbits are no longer pushed away from 
circularity and the turnover in the eccentricity occurs much later -- both features that are expected in true inspirals. 
The different trajectories also plunge at somewhat more different points. There is some excess growth in eccentricity 
close to plunge, but this is due to the failure of the GHK flux expressions \erf{GHKEdot}--\erf{GHKLdot}. 
Using higher order expressions (see section~\ref{2PN}), eliminates this feature. We conclude that it does not make 
too much difference how the fix is imposed, but imposing the fix gives a significant improvement in the qualitative 
behaviour of the inspirals deep in the strong-field region of the spacetime.

In the limit of a nearly polar orbit, we find that $e\,(\partial e/\partial L_{z})|_{E,\iota}$ and 
$e\,(\partial e/\partial \iota)|_{E,L}$ both diverge like $1/L_{z}$, while $e\,(\partial e/\partial E)|_{L,\iota}$ 
is finite (this will be discussed in more detail in the next section). As a result, if we specify $\dot{L}_{z}$ 
and $\dot{\iota}$ and use \erf{circcond} to correct the circular piece of $\dot{E}$, the 
correction diverges in the limit $\iota \rightarrow \pi/2$. Therefore, when evolving $\iota$ one should use 
$\dot{E}$ and $\dot{\iota}$ to correct $\dot{L}_{z}$. However, if we evolve $Q$ instead of 
$\iota$, the functions $e\,(\partial e/\partial L_{z})|_{E,Q}$ and $e\,(\partial e/\partial Q)|_{E,L}$ are both 
finite at the pole and there is no problem. However, $e\,(\partial e/\partial L_z)|_{E,Q}$ vanishes for certain 
retrograde orbits. Using $\dot{E}$ and $\dot{Q}$ to correct $\dot{L}_z$ would therefore lead to a divergence at this point 
(unless $\dot{E}$ and $\dot{Q}$ are constrained to cancel appropriately). Thus, when evolving $Q$, it is better to use 
$\dot{L}_z$ and $\dot{Q}$ to correct $\dot{E}$. The modified $\dot{E}$ is
\begin{eqnarray}
\dot{E} &=& (1-e^2)^{\frac{3}{2}} \left [ (1-e^2)^{-\frac{3}{2}}\left(\dot{E}\right)_{GHK} (p,\iota,e,a) 
- \left(\dot{E}\right)_{GHK} (p,\iota,0,a) \right. \nonumber \\ && \left. 
- \frac{N_4(p,\iota)}{N_1(p,\iota)}\,\left(\dot{L}_z \right)_{GHK} (p,\iota,0,a) 
-\frac{N_5(p,\iota)}{N_1(p,\iota)} \, \left(\dot{Q}\right)_{GHK} (p,\iota,0,a)\right ] \label{Edotcorr}
\end{eqnarray}
In this, $(\dot{E},\dot{L}_z, \dot{Q} )_{GHK}$  refer to the prescription of the energy and angular momentum fluxes being used, 
e.g., \erf{GHKLdot}, \erf{GHKidot} in this case. If the fluxes are modified (e.g., to the 2PN expressions given in section~\ref{2PN}) 
the correction is applied as above, with $(\dot{E})_{GHK}$ replaced by $(\dot{E})_{2PN}$ etc.

%%%%%%%%%%%%%%%%%%%%%%%%%%%%%%%%%%%%%%%%%%%%%%%%%%%%%%%%%%%%%%%%%%%%%%%%%%%%%

\section{Correcting the behaviour of nearly polar inspirals}
\label{highinc}
\begin{figure}
\centerline{\includegraphics[keepaspectratio=true,height=5in,angle=-90]{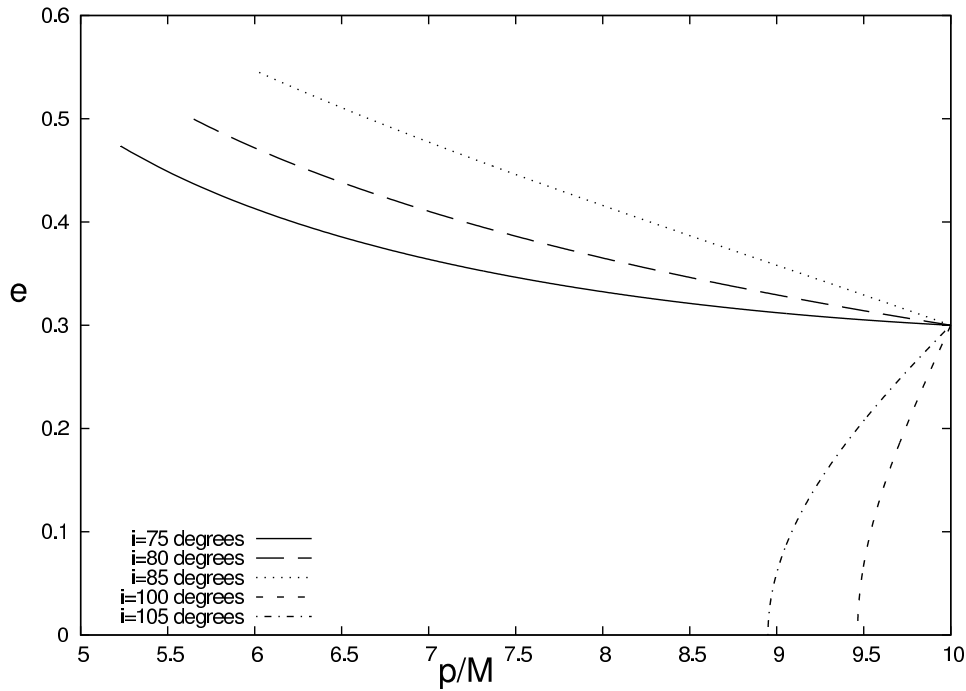}}
\caption{Inspirals into a black hole of spin $a=0.9\,{\rm M}$ for initial eccentricity $e=0.3$ and initial 
semi-latus rectum $p=10\,{\rm M}$. These inspirals were computed using the fluxes \erf{GHKEdot}--\erf{GHKLdot}, 
but without the small-eccentricity correction. The inspirals are shown for initial inclinations of 
$\iota=75^{\circ}$, $80^{\circ}$, $85^{\circ}$, $100^{\circ}$ and $105^{\circ}$.}
\label{KGpolar}
\end{figure}

\begin{figure}
\centerline{\includegraphics[keepaspectratio=true,height=5in,angle=-90]{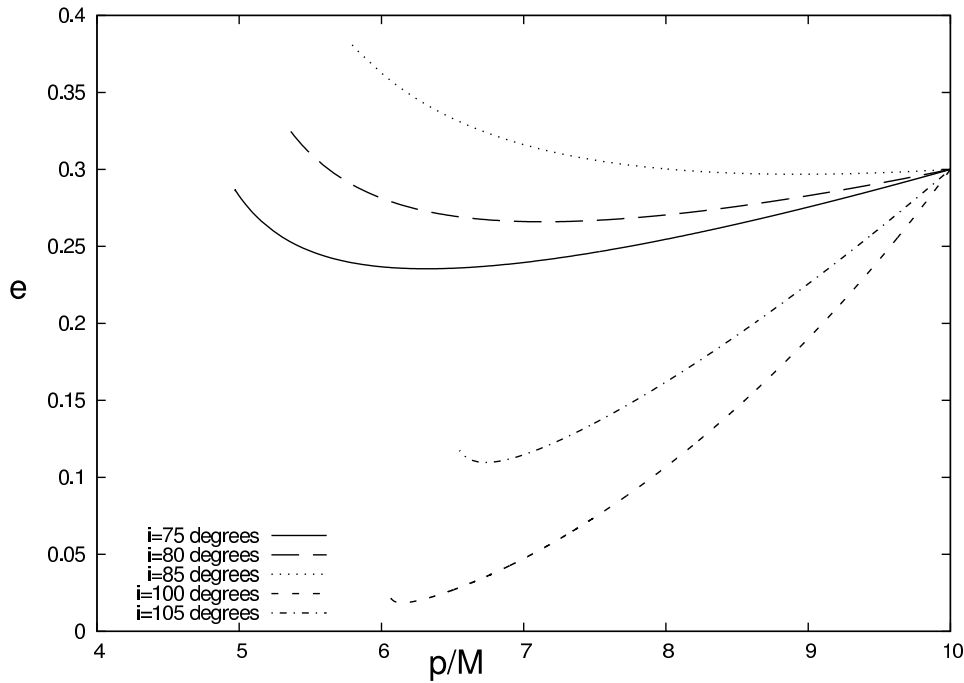}}
\caption{The same inspirals as in Figure~\ref{KGpolar}, but now computed using the fluxes \erf{GHKEdot}--\erf{GHKLdot} 
and the small-eccentricity correction \erf{Edotcorr}.}
\label{KGpolarpluscirc}
\end{figure}

The GHK hybrid scheme also exhibits some pathological behaviour for orbits that are nearly polar, i.e., with $\iota \approx \pi/2$. 
In Figure~\ref{KGpolar} we show inspirals for four different initial inclination angles, computed under the 
original GHK scheme. The inspirals have two strange properties. Firstly, the transition passing across the pole 
is not smooth -- there is a discontinuity between trajectories with $\iota$ slightly less than $90^\circ$, 
and those with $\iota$ slightly above $90^\circ$. In fact, it is not possible to compute trajectories in this approximation 
for $\iota$ slightly less than $90^{\circ}$, since it predicts an increase in $p$. Secondly, orbits with high prograde 
inclinations rapidly circularise, which is unphysical. This latter effect looks like a manifestation of the problem 
$\dot{e} \propto 1/e$ for small $e$ discussed in the previous section, but the coefficient of proportionality 
is now positive. We might hope that the correction for near-circular orbits discussed above will fix this problem. 
Figure~\ref{KGpolarpluscirc} shows the same set of inspirals, computed including the near-circular correction \erf{Edotcorr}. 
As expected, the correction does fix the rapid circularization problem, but the transition across the pole is 
still discontinuous.

The remaining problem near $\iota=\pi/2$ can again be understood by considering the behaviour of 
$\dot{e}$ (or $\dot{p}$) in that limit. The derivatives $\partial e/\partial L_{z}$ and $\partial e/\partial \iota$ 
are both singular, diverging like $\sec\iota$ as $\iota \rightarrow \pi/2$. This is clear from looking at $N_{2}(r)$ 
and $N_{3}(r)$ in that limit \erf{ederivs}. In fact, in terms of the Carter constant $Q$ of the polar orbit, we find
\begin{equation}
N_{2}(r) \approx \sqrt{Q} \, (-r^{2}+2\,M\,r-a^{2})\,\sec\iota, \qquad N_{3}(r) \approx Q\,(-r^{2}+2\,M\,r+a^{2})\,
\sec\iota \qquad \Rightarrow \qquad \frac{N_{3}(r)}{N_{2}(r)} \approx \sqrt{Q} 
\label{polarlim}
\end{equation}
The divergence of $\dot{e}$, which corresponds to a divergence in $\dot{r}_{p}$ and $\dot{r}_{a}$ 
is unphysical. It will be avoided if the singularities in the $L_{z}$ and $\iota$ fluxes cancel appropriately, 
which implies
\begin{equation}
\dot{L}_{z} \, \left(p,e,\iota=\frac{\pi}{2}\right) = -\sqrt{Q\left(p,e,\iota=\frac{\pi}{2}\right)}\,\dot{\iota} \,
\left(p,e,\iota=\frac{\pi}{2}\right) . \label{polarcond}
\end{equation}
There is nothing special about polar orbits which should allow us to specify an additional condition, so where does 
\erf{polarcond} come from? This is answered by looking at the derivative of the Carter constant
\begin{equation}
\dot{Q} = 2\,L_{z}\tan^{2}\iota\,\dot{L}_{z} + 2\,L_{z}^{2}\,\tan\iota\,\sec^2\iota\,\dot{\iota} = 2\,\sqrt{Q}\,
\frac{\sin\iota}{\cos\iota}\,\left(\dot{L}_{z} + \sqrt{Q}\,\frac{1}{\sin^{2}\iota}\,\dot{\iota}\right) \label{Qcond}
\end{equation}
The condition \erf{polarcond} is equivalent to requiring that $\dot{Q}$ is finite in the limit $\iota \rightarrow \pi/2$. 
The problem has arisen because we have chosen to evolve the inclination angle rather than evolve $Q$ directly. 
A real radiation field will generate a finite change in the Carter constant, and the rate of change of $\iota$ computed 
from this will automatically satisfy \erf{polarcond}. The main thing this tells us is that the constant inclination approximation, 
$\dot{\iota}=0$, is unphysical -- polar orbits lose angular momentum, which drives them to become retrograde. 
An alternative prescription for the evolution of $\iota$ is therefore required. If we specify $\dot{L}_z$ and 
$\dot{\iota}$ at some post-Newtonian order, the relation \erf{polarcond} will not hold exactly, as $\sqrt{Q}$ 
is accurate to arbitrary PN order. We can account for this by adding a correction to $\dot{L}_z$, as for the near 
circular correction. However, since this correction is only required at the pole, it is difficult to extend it smoothly to 
non-polar orbits. The need for a correction can be eliminated entirely if we evolve $Q$ instead of $\iota$, i.e., we choose 
to specify $\dot{Q}$ and derive $\dot{\iota}$ from this.

\begin{figure}
\centerline{\includegraphics[keepaspectratio=true,height=5in,angle=-90]{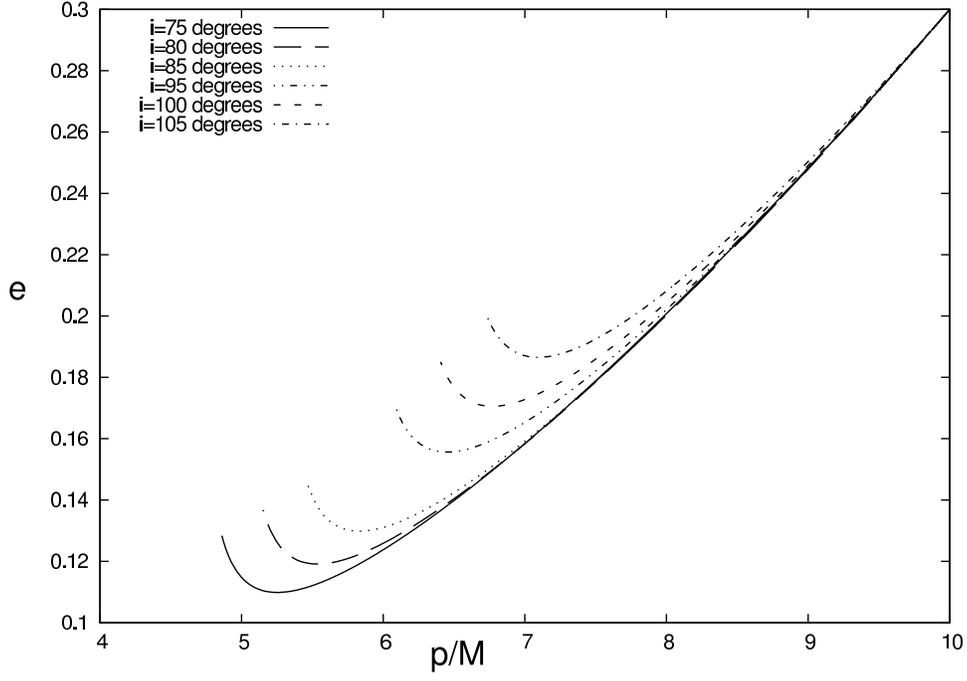}}
\caption{The same inspirals as in Figure~\ref{KGpolarpluscirc}, but now including evolution of the inclination angle using \erf{Qdot_Ryan}. We also show an inspiral with initial inclination of $\iota=95^{\circ}$ which could not be computed under the original version of the hybrid scheme.}
\label{AllFixPolar}
\end{figure}

Ryan's expression \cite{Ryan} for $\dot{Q}$, written in terms of our orbital elements is
\begin{equation}
\dot{Q}_{R} = -\frac{64}{5} \mu^3 \left (\frac{M}{p} \right )^3
(1-e^2)^{3/2} \sin^2\iota \left [ f_{3}(e) - \frac{a}{M}\,\left ( \frac{M}{p} \right)^{3/2} 
\cos\iota f_{6}(e) \right ] ,\
\label{Qdot_Ryan}
\end{equation} 
with
\begin{equation}
f_{3}(e) = 1 + \frac{7}{8}e^2 ,\qquad f_{6}(e) = \frac{85}{8} +
\frac{211}{8}e^2 + \frac{517}{64}e^4 .\
\end{equation}
We can use this in conjunction with the fluxes \erf{GHKEdot}--\erf{GHKLdot} to compute inspirals. 
Figure~\ref{AllFixPolar} shows the set of inspirals depicted in Figures~\ref{KGpolar} and \ref{KGpolarpluscirc}, 
but now evolved using \erf{Qdot_Ryan} instead of the constant inclination approximation. The resulting inspirals now change smoothly 
as the inclination angle is increased. However, the change in $\iota$ over the course of the inspiral is much too large when 
compared to the results of Teukolsky computations for circular orbits \cite{scott2}. This is the reason GHK chose to evolve 
$\iota$ rather than $Q$. In section~\ref{newQdot} we will show how to improve the approximation for $\dot{Q}$ to give more 
reasonable evolutions.

%%%%%%%%%%%%%%%%%%%%%%%%%%%%%%%%%%%%%%%%%%%%%%%%%%%%%%%%%%%%%%%%%%%%%%%%%%%%%

\section{Inclusion of higher order Post-Newtonian fluxes}
\label{2PN}
The corrections discussed in the previous two sections must be applied in any version of the hybrid scheme, and ensure that 
the inspiral properties are physically reasonable. However, the expressions used to evolve $E$, $L_z$ and $Q$ can be 
improved, to give better agreement with perturbative calculations \cite{scott,scott2,kgdk}.

\subsection{Equatorial orbits}
\begin{figure}
\centerline{\includegraphics[keepaspectratio=true,height=5in,angle=-90]{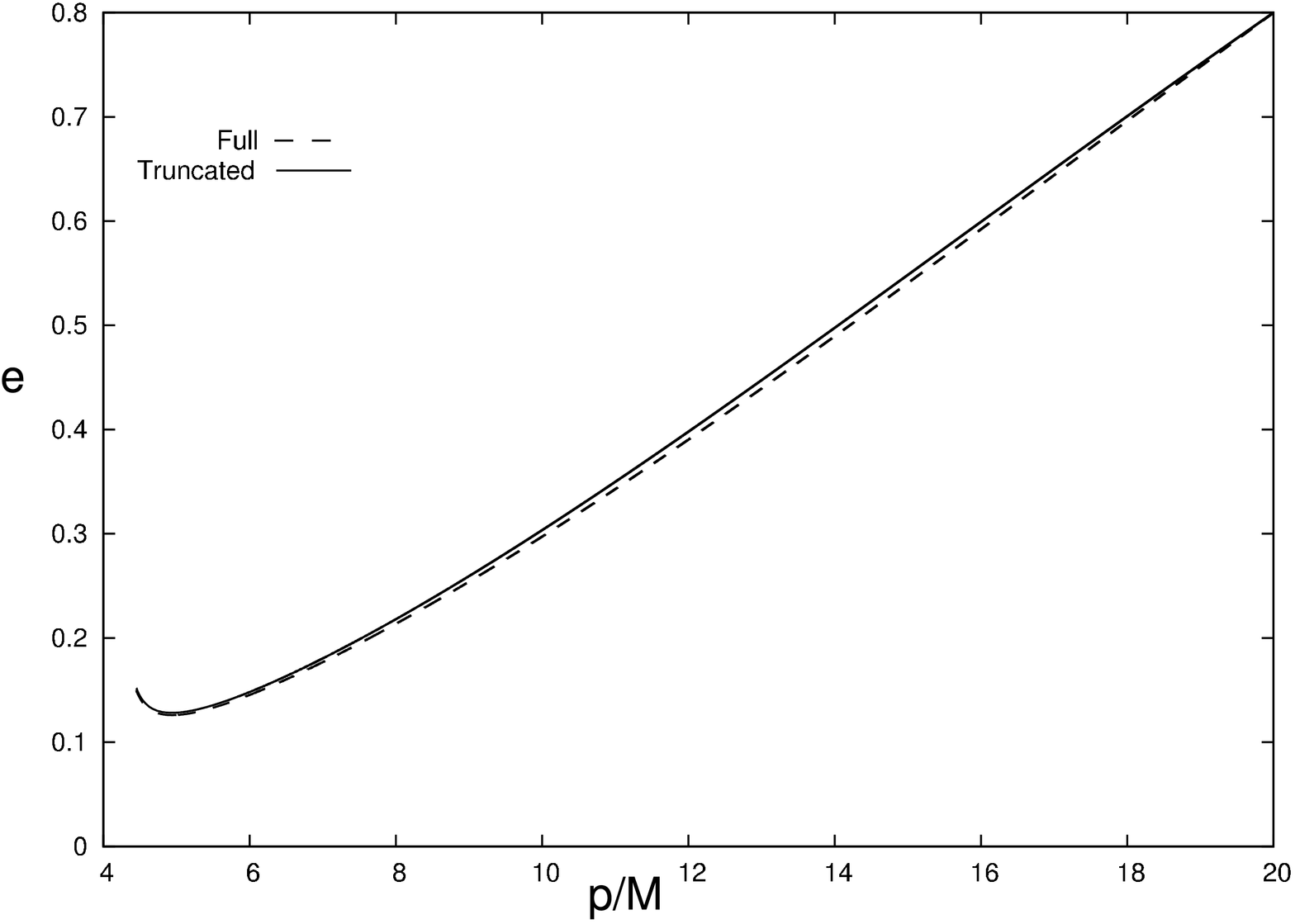}}
\caption{Approximate equatorial inspirals for an orbit with initial parameters
$p=20 M, e=0.8$. The dashed curve was produced by use of the full-$e$ Ryan fluxes while for the solid curve only 
${\cal O}(e^2)$ terms were retained. The black hole spin is $a=0.5 M$.}
\label{fig1}
\end{figure}

A natural way to extend GHK is to replace \erf{GHKEdot}--\erf{GHKLdot} with higher order expressions for the evolution 
of $E$ and $L_{z}$. Tagoshi \cite{tagoshi} derived 2.5 Post-Newtonian (PN) fluxes for the case of {\it equatorial} and 
eccentric orbits under 
the assumption of {\it small} eccentricity (Tagoshi  kept up to ${\cal O}(e^2) $ terms). The assumption of small $e$ 
is not as restrictive
as it initially sounds, as can be demonstrated by comparing inspirals generated using the GHK fluxes \erf{GHKEdot}--\erf{GHKLdot} 
to those generated using GHK truncated at ${\cal O}(e^2) $. This comparison is shown in Figure~\ref{fig1} and indicates that 
inspirals can be faithfully reproduced using the $e$-truncated fluxes provided $e \lesssim 0.8$. We expect the same to be true 
for the higher order PN fluxes that we construct below.

In his calculation, Tagoshi \cite{tagoshi} describes the test-body's orbital motion with the parameters 
$r_{0} $, $e_{j}$ and $v^2= M/r_{0}$. The parameter $r_{0} $ is the radius at which the potential, $V_{r}$, 
governing the radial motion has a minimum. Then, $e_j$ is defined by the requirement that $r= r_{0} (1+e_j)$ 
is a turning point for the radial motion, i.e.
$V_r(r=r_{0} (1+e_j))=0$. It turns out that the radial motion $r(t)$, to ${\cal O}(e_{j}^2)$ accuracy, is given by,
\begin{equation}
r(t)= r_{0}  [ 1 + e_j r^{(1)}(t) + e_{j}^2 r^{(2)}(t) + 
{\cal O}(e_{j}^3) ]   ,\
\label{small_e_motion}
\end{equation}
where
\begin{eqnarray}
r^{(1)}(t) &=& \cos\Omega_r t  ,\  \\
r^{(2)}(t) &=& q_3 (1-\cos\Omega_r t) + q_4 (1-\cos2\Omega_r t) .\
\end{eqnarray}
The quantity $\Omega_r$ is the angular frequency associated with
the radial motion, while $q_3,q_4$ are functions of $r_{0} $
and the hole's spin $q= a/M$ (we write explicitly only the former),
\begin{equation}
q_3(r_{0}) = 1 -\frac{M}{r_{0} } + 2q \left(\frac{M}{r_{0} }\right)^{3/2} -
(6+q^2)\left(\frac{M}{r_{0} }\right)^{2} 
+ 20q\left(\frac{M}{r_{0} }\right)^{5/2} + {\cal O}(v^6) .\
\label{param_q3} 
\end{equation}
From equation~\erf{small_e_motion} we can immediately see that the 
apoastron and periastron are given by,
\begin{eqnarray}
r_a &=& r_{0}  (1+ e_j) + {\cal O}(e_{j}^3) ,\ 
\label{tag_ra} \\
r_p &=& r_{0} ( 1 - e_j + 2 q_3 e_{j}^2 ) + {\cal O}(e_{j}^3) .\
\label{tag_rp}
\end{eqnarray}
The averaged fluxes at infinity are given by \cite{tagoshi},
\begin{eqnarray}
\dot{E}&=& -\frac{32}{5}\frac{\mu^2}{M^2} v^{10} \left[ 1 - \frac{1247}{336}v^2
+ 4\pi v^3 - \frac{73}{12}qv^3 -\frac{44711}{9072} v^4 + \frac{33}{16}q^2 v^4 \right.
\nonumber \\
&& \left. -\frac{8191}{672}\pi v^5 + \frac{3749}{336}q v^5 + e_{j}^2 \left( \frac{37}{24}
-\frac{65}{21}v^2 + \frac{1087}{48}\pi v^3 -\frac{211}{6}q v^3 \right. \right.
\nonumber \\
&&  \left. \left. -\frac{465337}{9072}v^4 + \frac{105}{8}q^2 v^4 
-\frac{118607}{1344}\pi v^5 -\frac{95663}{672}q v^5 \right ) 
+ {\cal O}(v^6) \right] ,\
\nonumber \\
\nonumber \\
\dot{L}_z &=& -\frac{32}{5}\frac{\mu^2}{M} v^{7} \left[ 1 - \frac{1247}{336}v^2
+ \left(4\pi -\frac{61}{12}q\right) v^3 +  \left(-\frac{44711}{9072}  
+ \frac{33}{16}q^2 \right)v^4 \right.
\nonumber \\
&& \left. + \left(-\frac{8191}{672}\pi  + \frac{417}{56}q\right) v^5 
+ e_{j}^2 \left( -\frac{5}{8} +\frac{749}{96}v^2 + \left(\frac{49}{8}\pi  
-\frac{57}{4}q\right) v^3 \right. \right.
\nonumber \\
&&  \left. \left. + \left(-\frac{232181}{6048} + \frac{203}{32}q^2\right) v^4 
+ \left(\frac{773}{336}\pi  -\frac{28807}{224}q \right) v^5 \right ) 
+ {\cal O}(v^6)  \right] .\
\label{tag_flux}
\end{eqnarray}

\begin{figure}
\centerline{\includegraphics[keepaspectratio=true,height=5in,angle=-90]{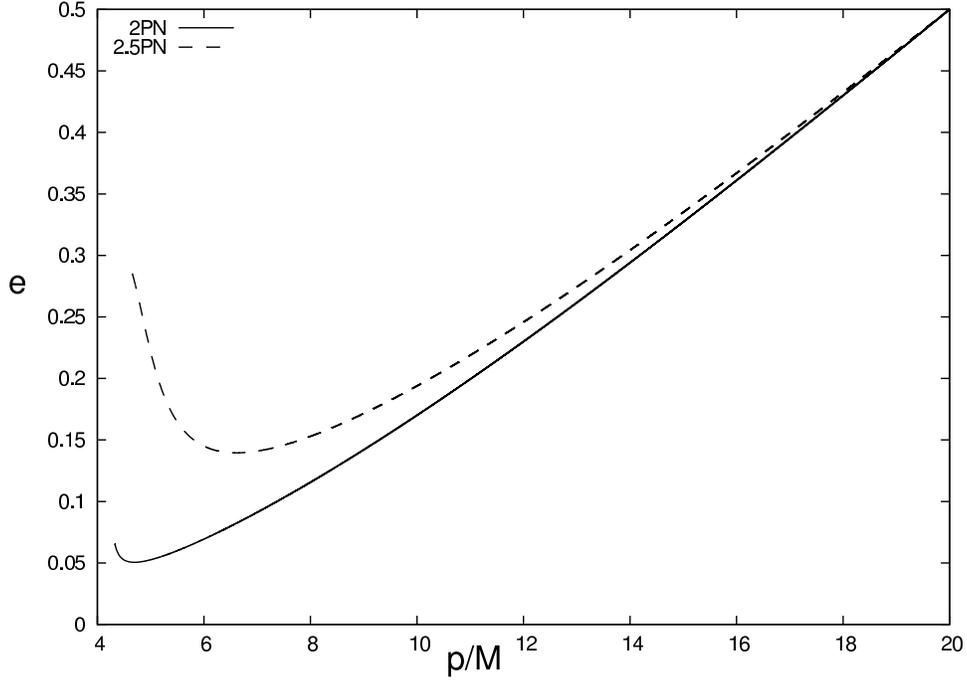}}
\caption{Determining which PN order is more reliable: the figure shows inspirals with initial parameters $p=20M, e=0.5$ and spin $a= 0.5M$ generated using the fluxes (\ref{new_Edot}), (\ref{new_Ldot}) truncated at 2.5PN (dashed curve) and 2PN (solid curve). It is clear that 2PN is the best choice.}
\label{fig2}
\end{figure}

In order to adapt these fluxes to the GHK scheme we first need to rewrite them in terms of our parameters 
$p$ and $e$. Assuming a small
eccentricity, $e \ll 1$, we have,
\begin{eqnarray}
r_a &=& \frac{p}{1-e}= p( 1 + e + e^2 ) + {\cal O}(e^3) ,\
\\
r_p &=& \frac{p}{1+e} = p(1-e + e^2 ) + {\cal O}(e^3) .\
\end{eqnarray}
Combining these with equations~\erf{tag_ra}--\erf{tag_rp} leads to
\begin{eqnarray} 
e_j &=& e + e^2 q_3(p) ,\
\\
\nonumber \\
\frac{1}{r_{0}} &=& \frac{1}{p}\left [ 1 + e^2 (q_3(p) -1) \right] .\
\end{eqnarray}
It is now straightforward now to rewrite the fluxes (\ref{tag_flux}) in terms of $p$ and $e$. To enhance accuracy, 
we adopt the approach `where possible, include higher order terms in $e$' (for example, by comparing with Ryan's expressions). 
In particular, we must have the factor $(1-e^{2})^{\frac{3}{2}}$ to ensure the behaviour is qualitatively correct for 
high eccentricity, as discussed earlier in the context of the near-circular correction. We find,
\begin{eqnarray}
\dot{E}&=& -\frac{32}{5} \frac{\mu^2}{M^2} \left(\frac{M}{p}\right)^5 
(1-e^2)^{3/2}\left [ g_1(e) -q\left(\frac{M}{p}\right)^{3/2} g_2(e) 
-\left(\frac{M}{p}\right) g_3(e) +
\pi\left(\frac{M}{p}\right)^{3/2} g_4(e) -\left(\frac{M}{p}\right)^2 g_5(e) \right.
\nonumber \\
&& \left. + q^2\left(\frac{M}{p}\right)^2 g_6(e) -
\pi\left(\frac{M}{p}\right)^{5/2} g_7(e)
+ q\left(\frac{M}{p}\right)^{5/2} g_8(e) \right ] ,\
\label{new_Edot}
\nonumber \\
\\
\dot{L}_z&=& -\frac{32}{5} \frac{\mu^2}{M} \left(\frac{M}{p}\right)^{7/2} 
(1-e^2)^{3/2}
\left [ g_9(e) -q\left(\frac{M}{p}\right)^{3/2} g_{10}(e) 
-\left(\frac{M}{p}\right) g_{11}(e) +
\pi\left(\frac{M}{p}\right)^{3/2} g_{12}(e) 
-\left(\frac{M}{p}\right)^2 g_{13}(e) \right.
\nonumber \\
&& \left. + q^2\left(\frac{M}{p}\right)^2 g_{14}(e) 
-\pi\left(\frac{M}{p}\right)^{5/2} g_{15}(e)
+ q\left(\frac{M}{p}\right)^{5/2} g_{16}(e) \right ] ,\
\label{new_Ldot}
\end{eqnarray}
where the various $e$-dependent coefficients are,
\begin{eqnarray}
g_1(e) &=& 1 + \frac{73}{24} e^2  + \frac{37}{96}e^4, \quad\quad\quad  
g_2(e)= \frac{73}{12} + \frac{823}{24} e^2 + \frac{949}{32}e^4
+ \frac{491}{192}e^6 ,\  
\nonumber \\
\nonumber \\
g_3(e) &=& \frac{1247}{336} + \frac{9181}{672} e^2 ,\quad\quad 
g_4(e)  = 4 + \frac{1375}{48} e^2  ,\
\nonumber \\
\nonumber \\
g_5(e) &=& \frac{44711}{9072} + \frac{172157}{2592} e^2 ,
\quad\quad 
g_6(e) = \frac{33}{16} + \frac{359}{32} e^2  ,\
\nonumber \\
\nonumber \\
g_7(e) &=& \frac{8191}{672} + \frac{44531}{336} e^2   ,
\quad\quad\quad 
g_8(e) = \frac{3749}{336} - \frac{5143}{168} e^2    ,\
\nonumber \\
\nonumber \\
g_9(e) &=& 1 + \frac{7}{8} e^2    ,
\quad\quad\quad\quad\quad\quad
g_{10}(e) = \frac{61}{12} + \frac{119}{8} e^2 +\frac{183}{32}e^4  ,\ 
\nonumber \\
\nonumber \\
g_{11}(e) &=& \frac{1247}{336} + \frac{425}{336} e^2  ,\quad\quad\quad\quad
g_{12}(e) = 4 + \frac{97}{8} e^2 ,\
\nonumber \\
\nonumber \\
g_{13}(e) &=& \frac{44711}{9072} + \frac{302893}{6048} e^2 ,\quad\quad 
g_{14}(e) = \frac{33}{16} + \frac{95}{16} e^2 ,\
\nonumber \\
\nonumber \\
g_{15}(e) &=& \frac{8191}{672} + \frac{48361}{1344} e^2 ,\quad\quad\quad
g_{16}(e) = \frac{417}{56} - \frac{37241}{672} e^2 .\
\end{eqnarray}
It is a well known fact that PN expansions are characterised by poor convergence, that is, a higher PN order does not 
necessarily mean more accurate result. The same behaviour is found in these fluxes too. After some testing we found that 
the optimal order is 2PN (in agreement with the literature on the subject). This is illustrated in Figure~\ref{fig2} 
where we compare 2PN with 2.5PN fluxes. Note that in this and subsequent figures, we are including the near-circular 
correction \erf{Edotcorr} and, for inclined inspirals, are evolving the inclination by prescribing $\dot{Q}$.

\begin{figure}
\centerline{\includegraphics[keepaspectratio=true,height=5in,angle=-90]{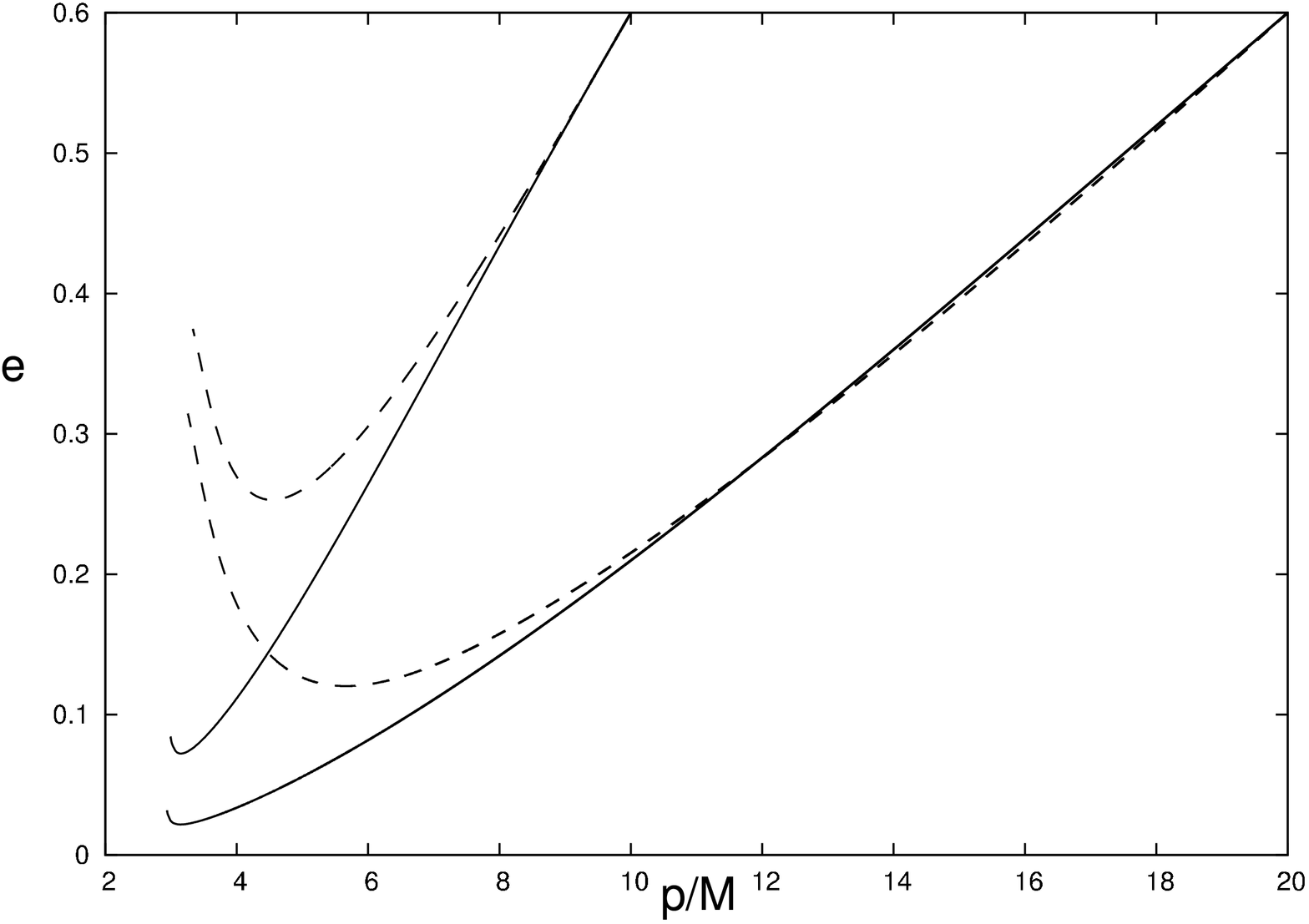}}
\caption{Equatorial inspirals generated with the help of the new fluxes (\ref{new_Edot}), (\ref{new_Ldot}) (solid curves) compared to the original GHK inspirals (dashed curves).
The initial parameters are $p=20M,e=0.6$ and $p=10 M, e=0.6$ for black hole spin $a=0.8M$.}
\label{fig4}
\end{figure}

Figure~\ref{fig4} illustrates a set of inspirals computed using the revised fluxes \erf{new_Edot}--\erf{new_Ldot} 
(truncated at 2PN level). These represent equatorial inspirals into a black hole of spin $a=0.8 M$. This is one case 
for which the original GHK scheme did not perform accurately. The revised inspirals are a significant improvement over the 
GHK results, particularly near the separatrix. Comparison of the numerical values of $\dot{L}_z$ and $\dot{E}$ with the results 
of Teukolsky-based calculations \cite{kgdk} also indicate we are making a significant improvement.

%%%%%%%%%%%%%%%%%%%%%%%%%%%%%%%%%%%%%%%%%%%%%%%%%%%%%%%%%%%%%%%%%%%%

\subsection{Extension to non-equatorial inspirals}
The results presented in the previous section apply only to equatorial orbits, but ultimately the goal is to improve 
the inspiral of generic (inclined and eccentric) orbits. The only available higher order PN expansion for inclined orbits 
has been derived by Shibata \etal \cite{shibata} for the case of {\it circular} orbits with {\it small} inclination.  
Shibata \etal use the inclination parameter
\begin{equation}
y = \frac{Q}{L_{z}^2 + a^2(1-E^2)} ,\
\end{equation}  
and assumed that this is a small number. 
The relation between $y$ and our inclination angle $\iota$ is,
\begin{equation}
\cos^2\iota = \frac{1}{1+ y[1+ a^2(1-E^2)/L_{z}^2]} ,\
\end{equation}
which for small $\iota,y$ and to 2PN accuracy becomes,
\begin{equation}
y \approx \iota^2 [ 1 -a^2(1-E^2)/L_{z}^2 ] = \iota^2 [1 -(a/M)^2 (M/p)^2 ] .\
\end{equation}
The fluxes derived by Shibata et. al. are given by,
\begin{eqnarray}
\dot{E} &=& -\frac{32}{5}\frac{\mu^2}{M^2} v^{10}
\left [ 1  -\frac{1247}{336} v^2  + 4\pi v^3
-\frac{73}{12}q v^3 (1-y/2)  \right.
\nonumber \\ 
&& \left. -\frac{44711}{9072}v^4 + \frac{33}{16}q^2 v^4 
-\frac{527}{96}q^2 v^4 y + {\cal O}(v^5) \right ] ,\
\nonumber \\
\nonumber \\
\dot{L}_{z} &=& -\frac{32}{5}\frac{\mu^2}{M} v^{7} \left [
1 -y/2 -\frac{1247}{336}v^2(1-y/2) + 4\pi v^3 (1-y/2) 
-\frac{61}{12}q v^3 (1-3y/2)  \right.
\nonumber \\
&& \left. -\frac{44711}{9072}v^4 (1-y/2)
+ q^2 v^4 \left (\frac{33}{16} -\frac{229}{32}y \right ) + {\cal O}(v^5) \right ].\
\label{shib_flux}
\end{eqnarray}
At 2PN accuracy, we find that $ 1-y/2 = 1- \iota^2/2 + \iota^{2}/2 (a/M)^2 (M/p)^2 $. Term by term comparison 
(where possible) between expressions (\ref{shib_flux}) and Ryan's fluxes \erf{GHKEdot}--\erf{GHKLdot} suggests the 
correspondence $ 1-y/2 \rightarrow \cos\iota $ and $ 1-3y/2 \rightarrow -1/2 + 3/2 \cos^2 \iota $ when one goes from small 
to arbitrary inclination for these terms. For the 2PN $q^2$ terms, we expect $\dot{E}$ to contain terms proportional to $1$ 
and $\cos^2\iota$, while $\dot{L}_z$ should contain pieces proportional to $\cos\iota$ and $\cos^3\iota$ \cite{poisson98}. 
This is born out by fits to circular Teukolsky computations \cite{scott} (see section~\ref{Teukfit}).

The cosine terms contribute in the equatorial plane and therefore, using \erf{new_Edot}--\erf{new_Ldot}, we know the 
$e$-dependent factors that multiply them. However, this is not true for the sine terms. Consequently, we expect terms 
that contain $\sin\iota$ to be incomplete and of modest accuracy. Nevertheless we have included them in our fluxes 
(and verified that for orbits not close to the equator these terms play an important role). Putting together all of the above, 
we extend the fluxes \erf{new_Edot}--\erf{new_Ldot} to generic orbits as follows
\begin{eqnarray}
(\dot{E})_{\rm 2PN} &=& -\frac{32}{5} \frac{\mu^2}{M^2} \left(\frac{M}{p}\right)^5 
(1-e^2)^{3/2}\left [ g_1(e) -q\left(\frac{M}{p}\right)^{3/2} g_2(e)
\cos\iota -\left(\frac{M}{p}\right) g_3(e) +
\pi\left(\frac{M}{p}\right)^{3/2} g_4(e)  \right.
\nonumber \\
&& \left. -\left(\frac{M}{p}\right)^2 g_5(e) 
+  q^2\left(\frac{M}{p}\right)^2 g_6(e)
 -\frac{527}{96} q^2 \left( \frac{M}{p}\right)^2 \sin^2\iota
+ {\cal O}(v^5)   \right ] ,\
\label{new_Edot_2}
\nonumber \\
\\
(\dot{L}_z)_{2PN} &=& -\frac{32}{5} \frac{\mu^2}{M} \left(\frac{M}{p}\right)^{7/2}
(1-e^2)^{3/2}
\left [ g_9(e)\cos\iota  + q\left(\frac{M}{p}\right)^{3/2} \{g_{10}^{a}(e) 
-\cos^2\iota g_{10}^{b}(e) \} 
-\left(\frac{M}{p}\right) g_{11}(e) \cos\iota 
\right.
\nonumber \\
&& \left. + \pi\left(\frac{M}{p}\right)^{3/2} g_{12}(e)\cos\iota 
-\left(\frac{M}{p}\right)^2 g_{13}(e) \cos\iota  
+ q^2\left(\frac{M}{p}\right)^2 \,\cos\iota \left(g_{14}(e) - \frac{45}{8}\,\sin^2\iota\right)   + {\cal O}(v^5) \right ] ,\
\label{new_Ldot_2}
\end{eqnarray}
where 
\begin{equation}
g_{10}^{a}(e)= \frac{61}{24} + \frac{63}{8}e^2   + \frac{95}{64}e^4 ,\ 
\quad\quad 
g_{10}^{b}(e)= \frac{61}{8} + \frac{91}{4}e^2  + \frac{461}{64}e^4 .\
\end{equation} 
As discussed earlier, we have included higher order eccentricity terms where possible and truncated the series at 2PN level. 
We must also remember that we do not actually use the circular pieces of the $\dot{E}$ expression \erf{new_Edot_2}, 
but determine these using the circular fix \erf{Edotcorr}.

\begin{figure}
\centerline{\includegraphics[keepaspectratio=true,height=5in,angle=-90]{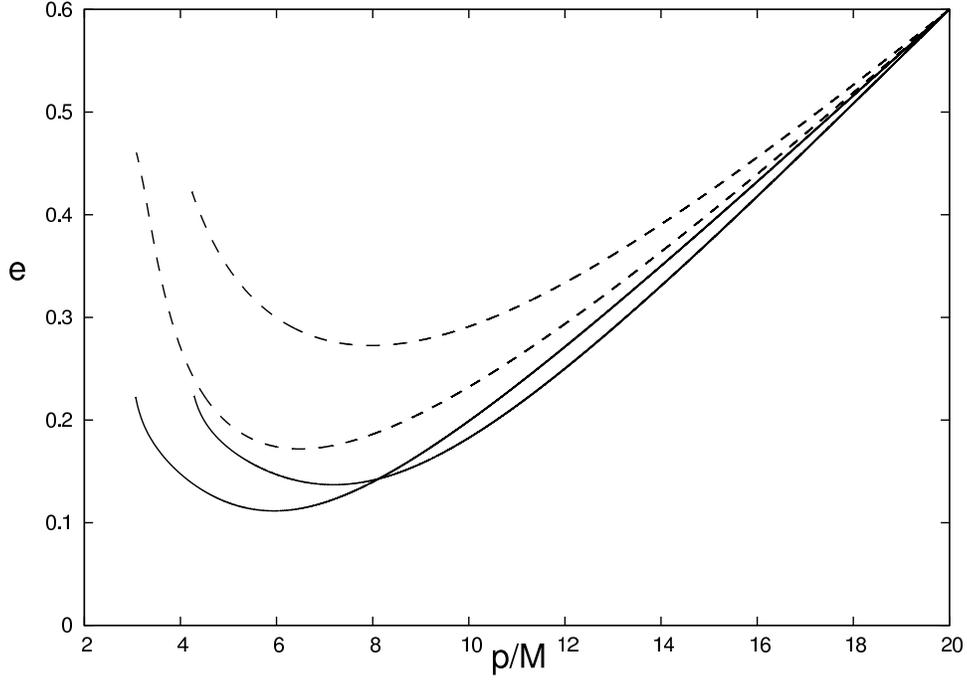}}
\caption{Inspirals generated with the help of the new fluxes 
(\ref{new_Edot_2}), (\ref{new_Ldot_2}) (solid lines) compared to the original GHK inspirals (dashed lines). The initial parameters are $p=20{\rm M},e=0.6, \iota = 30^\circ$ and $\iota= 60^\circ$ for black hole spin $a=0.9{\rm M}$.}
\label{fig8}
\end{figure}

Some numerical inspirals produced by incorporating these fluxes in the GHK scheme are shown in Figure~\ref{fig8}, along with 
the original GHK inspirals for comparison. These are inspirals into a black hole with spin $a=0.9 {\rm M}$ for two different 
inclinations -- $\iota=30^\circ$ and $\iota=60^\circ$. We are evolving $Q$ using Ryan's expression \erf{Qdot_Ryan}. 
As for the case of equatorial orbits, we find that the new fluxes produce inspirals that are more physically reasonable 
than the original GHK results.

%%%%%%%%%%%%%%%%%%%%%%%%%%%%%%%%%%%%%%%%%%%%%%%%%%%%%%%%%%%%%%%%%%%%%%%
\section{An improved prescription for the Carter constant flux}
\label{newQdot}
The constant inclination approximation \erf{GHKidot} is a ``spherical'' approximation in the sense that it is exact in the limit 
$a\rightarrow 0$. In section~\ref{highinc} we saw that $\dot{\iota}=0$ cannot be used to evolve nearly polar orbits. 
High inclination orbits offer the orbiting body an opportunity to more effectively probe the aspherical nature 
of the Kerr metric, thus it is not surprising that the ``spherical'' rule fails in such cases. Unfortunately, 
the available (and approximate) expressions for computing $\dot{Q}$ for generic orbits are extremely limited. 
Ryan's weak-field expression \cite{Ryan} gives rather poor results. As discussed in \cite{GHK}, the bulk of the 
change in $Q$ is reliably captured by equation~\erf{GHKidot}. Hence, we will attempt to add a correction to this expression 
in order to improve the inspirals.

%%%%%%%%%%%%%%%%%%%%%%%%%%%%%%%%%%%%%%%%%%%%%%%%%%%%%%%%%%%%%%%%%%%%%%%%%%%

\subsection{What is the connection between the ``spherical'' $\dot{Q}$ and the Kennefick-Ori formula ?}
Kennefick and Ori \cite{dan} give an expression for the evolution of the Carter constant in terms of the theta component 
of the self-force. Expressing the Carter constant as
\begin{equation}
Q = p_{\theta}^2 + \cos^2\theta [ a^2(\mu^2 -E^2) + \sin^{-2}\theta~ L^2_z ]
\equiv p_{\theta}^2 + H(E,L_z,\theta)  ,\
\end{equation}
the Kennefick-Ori formula (equation (11) of \cite{dan}) gives,
\begin{equation}
\dot{Q} = H_{,E} \dot{E} + H_{,L_z} \dot{L_z} + \frac{2\Sigma u^{\theta}}
{u^{t}}F_\theta  ,\
\label{ko1}
\end{equation}
where $ \Sigma = r^2 + a^2 \cos^2\theta$. Writing this explicitly,
\begin{equation}
\dot{Q}  = -2a^2 E \cos^2\theta~ \dot{E} + 2L_z \cot^2 \theta~ \dot{L}_z + \frac{2\Sigma u^\theta}{u^t} F_\theta .\
\label{ko2}
\end{equation}
This formula and the spherical $\dot{Q}$ \erf{GHKidot} have both been discussed and used in calculations in the extreme mass 
ratio literature over recent years \cite{GHK,scott,dan}. However, as yet there has been no discussion on their relation. 

In the $a\rightarrow 0$ limit, the first term of \erf{ko2} vanishes, $\Sigma \rightarrow r^{2}$, the Carter constant becomes 
$Q=L_{\rm tot}^{2}-L_{z}^{2}$ (where $L_{\rm tot}$ is the total angular momentum) and the inclination angle $\iota$ becomes the true inclination 
of the orbital plane, $\tan^2\iota=Q/L_{z}^{2}$. Taking the y-axis to lie in the orbital plane, without loss of generality, the Boyer-Lindquist 
coordinates of the particle at any point of the orbit obey the relation
\begin{equation}
f(\theta,\phi) = \cos\theta - \sin\theta\,\cos\phi\,\tan\iota = 0 \;.
\end{equation}
Symmetry ensures that the self-force has no component perpendicular to the orbital plane and therefore
\begin{equation}
{\bf u}\,\cdot\,{\bf n} = 0 = {\bf F}\,\cdot\,{\bf n} \qquad \Rightarrow \qquad \frac{F^{\theta}}{F^{\phi}} = 
\frac{u^{\theta}}{u^{\phi}} = \frac{\sec\phi\,\tan\phi}{\sec^{2}\theta \,\tan\iota}=\sin^{2}\theta \,
\sqrt{\tan^{2}\iota-\cot^{2}\theta} \;.
\end{equation}
In this, $n_{\alpha} = \partial f/\partial x^{\alpha}$ is the normal to the orbital plane and $u^{\alpha} = \rmd x^{\alpha}/\rmd \tau$ is the particle four-velocity. The third term in equation 
\erf{ko2} thus becomes
\begin{equation}
\frac{2\Sigma u^\theta}{u^t} F_\theta = \frac{2\,r^{4}\,u^{\theta}\,F^{\theta}}{u^{t}} = 
2\,\left(\frac{u^{\theta}}{u^{\phi}}\right)^2\,r^{4} \, \frac{u^{\phi}\,F^{\phi}}{u^{t}} = 
2\,\frac{(\tan^{2}\iota-\cot^{2}\theta)}{u^{t}}\,u_{\phi}\,F_{\phi} \;.
\end{equation}
But, $u_{\phi}=L_{z}$ and $F_{\phi} = \rmd L_{z}/\rmd \tau$, so $F_{\phi}/u^{t} = \dot{L}_{z}$ and 
\erf{ko2} becomes
\begin{equation}
\dot{Q} = 2\,(\cot^{2}\theta+\tan^{2}\iota-\cot^{2}\theta)\,L_{z}\,\dot{L}_{z} = 2\,\tan^{2}\iota\,L_{z}\,
\dot{L}_{z}=2\,Q\,\frac{\dot{L}_{z}}{L_{z}} \;,
\end{equation}
which brings \erf{ko2} to the form \erf{GHKidot}. Switching back on the spin makes (\ref{ko2}) depart from its spherical value 
and consequently $\iota$ to change. It is easy to deduce that some leading order (with respect to the spin) terms will be 
provided by the third term in (\ref{ko2}). This is because the leading order change in $F_\theta$ is linear in $a$, as can 
be found from Ryan's expressions for the self-force \cite{Ryan}. The first term, on the other hand, contributes only at 
${\cal O}(a^2)$ and beyond.

%%%%%%%%%%%%%%%%%%%%%%%%%%%%%%%%%%%%%%%%%%%%%%%%%%%%%%%%%%%%%%%%%%%%%%%%%%%%%%%%%%%%%%%%%%%%%%%%%%%%%%%%%%%

\subsection{An improved formula for $\dot{Q}$}
If we expand the spherical formula \erf{GHKidot} at the PN order of Ryan's expression, we find
\begin{equation}
\dot{Q}_{sph} = -\frac{64}{5} \mu^3 \left (\frac{M}{p} \right )^3
(1-e^2)^{3/2} \sin^2\iota \left [ f_{3}(e) + \frac{q}{\cos\iota}
\left ( \frac{M}{p} \right)^{3/2} \{ f_{4}(e) -\cos^2\iota f_{6}(e) \} 
\right ] ,\
\label{Qdot_sph2}
\end{equation}
where $ f_{4}(e) = 61/24 + 63 e^2/ 8 + 95 e^4/64 $ as in equation \erf{GHKefactors}, and $q=a/M$ as before. We see that the ${\cal O}(a)$ piece of 
this expression is not the same as in equation (\ref{Qdot_Ryan}) which by construction  includes {\it all} terms of this order. 
We conclude that there is an ${\cal O}(a)$ piece {\it missing} in the spherical $\dot{Q}$ formula. Written explicitly,
\begin{equation}
\delta_{Q} \equiv \dot{Q}_{R} - \dot{Q}_{sph} = \frac{64}{5}\mu^3 
\left ( \frac{M}{p}  \right )^{9/2} (1-e^2)^{3/2} q\frac{\sin^2\iota}{\cos\iota}
f_{4}(e) .\ 
\end{equation}
This piece represents the leading ``aspherical'' contribution to $\dot{Q}$. This gives rise to an evolution in $\iota$. We note, 
however, that this term is {\it divergent} at the pole, since it is proportional to $1/\cos\iota$. This is a manifestation of 
the near polar problem discussed in section~\ref{highinc}. In reality, $\dot{Q}$ will be finite at the pole, so physically 
$\dot{\iota}$ must be what is required to cancel all of the divergent pieces of the ``spherical'' $\dot{Q}$. We can therefore 
derive an improved prescription for $\dot{Q}$ by expanding the spherical formula \erf{GHKidot} and removing all terms that 
diverge at the pole. In this way, we derive from the 2PN angular momentum flux \erf{new_Ldot_2} a new expression for $\dot{Q}$ of the form
\begin{eqnarray}
\dot{Q} &=& \frac{2\,\sqrt{Q}\,\sin\iota}{\cos\iota}\, \left(\frac{\rmd L_z}{\rmd t}\right)_{2PN} - (\mbox{terms 
proportional to $\sec\iota$}) \nonumber \\
&=& -\frac{64}{5} \frac{\mu^2}{M} \left(\frac{M}{p}\right)^{7/2} \,\sqrt{Q}\,\sin\iota\,
(1-e^2)^{3/2}\,\left[ g_9(e) - q\left(\frac{M}{p}\right)^{3/2} \cos\iota g_{10}^{b}(e) -\left(\frac{M}{p}\right) g_{11}(e)
\right. \nonumber \\ && \left. + \pi\left(\frac{M}{p}\right)^{3/2} g_{12}(e) 
-\left(\frac{M}{p}\right)^2 g_{13}(e)
+ q^2\left(\frac{M}{p}\right)^2 \,\left(g_{14}(e) - \frac{45}{8}\,\sin^2\iota\right)  \right] \label{Qdot_new}
\end{eqnarray}
This method for prescribing $\dot{Q}$ removes the divergences at the pole, but is incomplete, since it tells us 
nothing about the pieces of $\dot \iota$ that vanish at the pole (or equivalently, the corrections to the non-divergent pieces 
of $\dot{Q}$ that come from $\dot{\iota} \neq 0$). Nonetheless, this is an improved prescription for the evolution of $Q$. 
In the next section we derive a further improvement to the circular piece of the $\dot{Q}$ expression.

%%%%%%%%%%%%%%%%%%%%%%%%%%%%%%%%%%%%%%%%%%%%%%%%%%%%%%%%%%%%%%%%%%%%%%%%%%%%%%%%%%%%%%%%%%%%%%%%%%%%%%%%%%%%%%%

\subsection{Fitting to Teukolsky data}
\subsubsection{Circular orbits}
\label{Teukfit}
The evolution of circular-inclined orbits has been computed accurately by solution of the Teukolsky equation \cite{scott,scott2}. 
We can use these results to improve the circular piece of our expression for $\dot{Q}$. It turns out that the best results are 
obtained by fitting functions to the Teukolsky $\dot{L}_z$ and $\dot{\iota}$ data, and deriving $\dot{Q}$ from these, 
via expression \erf{Qcond}.

Using data provided by Scott Hughes, we find that a good fit to the angular momentum flux $\dot{L}_{z}$ for circular orbits is given by the function
\begin{eqnarray}
(\dot{L}_z)_{\rm fit} &=& -\frac{32}{5} \frac{\mu^2}{M} \left(\frac{M}{p}\right)^{7/2} 
\left [ \cos\iota  + q\left(\frac{M}{p}\right)^{3/2} \left(\frac{61}{24} 
-\frac{61}{8}\,\cos^2\iota \right) -\frac{1247}{336}\,\left(\frac{M}{p}\right) \cos\iota 
 + 4\pi\,\left(\frac{M}{p}\right)^{3/2}\,\cos\iota \right.\nonumber \\ && \left.
-\frac{44711}{9072}\,\left(\frac{M}{p}\right)^2 \cos\iota  
+ q^2\left(\frac{M}{p}\right)^2 \,\cos\iota \left(\frac{33}{16} - \frac{45}{8}\,\sin^2\iota\right) 
 + \left(\frac{M}{p}\right)^{\frac{5}{2}} \left\{ q\, \left(d_1^a + d^b_1\,\left(\frac{M}{p}\right)^{\frac{1}{2}} + d^c_1\,
\left(\frac{M}{p}\right)\right) \right.\right.\nonumber \\ && \left. + q^3\,
\left(d_2^a + d^b_2\,\left(\frac{M}{p}\right)^{\frac{1}{2}} + d^c_2\,\left(\frac{M}{p}\right)\right)
+\cos\iota\left(c_1^a + c^b_1\,\left(\frac{M}{p}\right)^{\frac{1}{2}} + 
c^c_1\,\left(\frac{M}{p}\right) \right) \right. \nonumber \\ && \left. +q^2\,\cos\iota \,\left(c_2^a + c^b_2\,\left(\frac{M}{p}\right)^{\frac{1}{2}} + 
c^c_2\,\left(\frac{M}{p}\right)\right)
+ q^4\,\cos\iota\,\left(c_3^a + 
c^b_3\,\left(\frac{M}{p}\right)^{\frac{1}{2}}+ c^c_3\,\left(\frac{M}{p}\right) \right) \right.\nonumber \\ && \left. + 
q\,\cos^2\iota\left(c^a_4+ c^b_4\,\left(\frac{M}{p}\right)^{\frac{1}{2}}+ c^c_4\,\left(\frac{M}{p}\right)\right)
+q^3\,\cos^2\iota\,\left(c_5^a + c^b_5\,
\left(\frac{M}{p}\right)^{\frac{1}{2}} + c^c_5\,\left(\frac{M}{p}\right) \right) \right.\nonumber \\ && \left. + q^2\,\cos^3\iota\,
\left(c_6^a + c^b_6\,\left(\frac{M}{p}\right)^{\frac{1}{2}} + c^c_6\,
\left(\frac{M}{p}\right)\right)
+q^4\,\cos^3\iota\,\left(c_7^a + 
c^b_7\,\left(\frac{M}{p}\right)^{\frac{1}{2}} + c^c_7\,\left(\frac{M}{p}\right) \right) \right. \nonumber \\ && \left.\left.
+q^3\,\cos^4\iota\,\left(c_8^a + c^b_8\,
\left(\frac{M}{p}\right)^{\frac{1}{2}} + c^c_8\,\left(\frac{M}{p}\right) \right)
+q^4\,\cos^5\iota\,\left(c_9^a  
+ c^b_9\,\left(\frac{M}{p}\right)^{\frac{1}{2}} + c^c_9\,\left(\frac{M}{p}\right)\right) \right\} \right. \nonumber \\ && \left. + \left(\frac{M}{p}\right)^{\frac{7}{2}} \,q\,\cos\iota\,\left\{ f^a_1 + f^b_1\,\left(\frac{M}{p}\right)^{\frac{1}{2}} + q\,\left(f^a_2 + f^b_2\,\left(\frac{M}{p}\right)^{\frac{1}{2}}\right) + q^2\,\left(f^a_3 + f^b_3\,\left(\frac{M}{p}\right)^{\frac{1}{2}}\right) \right. \right. \nonumber \\ && \left. \left. + \cos^2\iota \,\left(f^a_4 + f^b_4\,\left(\frac{M}{p}\right)^{\frac{1}{2}}\right) + q\,\cos^2\iota \,\left(f^a_5 + f^b_5\,\left(\frac{M}{p}\right)^{\frac{1}{2}}\right) + q^2\,\cos^2\iota\,\left(f^a_6 + f^b_6\,\left(\frac{M}{p}\right)^{\frac{1}{2}}\right)\right\}
\right ] \nonumber \\ \mbox{where } && d^a_1 = -10.7420, \qquad d_1^b = 28.5942, \qquad 
d^c_1 = -9.07738,\qquad d^a_2 = -1.42836, \qquad d_2^b = 10.7003, \nonumber \\ && d^c_2 = -33.7090, \qquad c^a_1 = -28.1517, \qquad c^b_1 = 60.9607, \qquad c^c_1 = 40.9998, \qquad c^a_2 = -0.348161, \nonumber \\ 
&& c^b_2 = 2.37258, \qquad  c^c_2 = -66.6584, \qquad c^a_3 = -0.715392, \qquad c^b_3 = 3.21593, \qquad c^c_3 = 5.28888, 
\nonumber \\  && c^a_4 = -7.61034, \qquad  c^b_4 = 128.878, \qquad c^c_4 = -475.465, \qquad c^a_5 = 12.2908, \qquad c^b_5 = -113.125, \nonumber \\ && c^c_5 = 306.119, \qquad  c^a_6 = 40.9259, \qquad c^b_6 = -347.271, \qquad c^c_6 = 886.503, \qquad c^a_7 = -25.4831, \nonumber \\ 
&& c^b_7 = 224.227, \qquad  c^c_7 = -490.982, \qquad c^a_8 = -9.00634, \qquad c^b_8 = 91.1767, \qquad c^c_8 = -297.002, \nonumber \\ && c^a_9 = -0.645000, \qquad  c^b_9 = -5.13592, \qquad c^c_9 = 47.1982, \qquad f^a_1 = -283.955, \qquad f^b_1 = 736.209, \nonumber \\ && f^a_2 = 483.266, \qquad f^b_2 = -1325.19, \qquad f^a_3 = -219.224, \qquad f^b_3 = 634.499, \qquad f^a_4 = -25.8203, \nonumber \\ && f^b_4 = 82.0780, \qquad f^a_5 = 301.478, \qquad f^b_5 = -904.161, \qquad f^a_6 = -271.966, \qquad f^b_6 = 827.319. 
\label{Ldotfit}
\end{eqnarray}

Similarly, a good fit to the evolution of $\iota$ is given by
\begin{eqnarray}
(\dot{\iota})_{\rm fit}  &=& \frac{32}{5} \frac{\mu^2}{M} \,q\,\frac{\sin^2\iota}{\sqrt{Q}}\,\left(\frac{M}{p}\right)^{5} 
\left[ \frac{61}{24} + \left(\frac{M}{p}\right)\left(d_1^a + d^b_1\,\left(\frac{M}{p}\right)^{\frac{1}{2}} + 
d^c_1\,\left(\frac{M}{p}\right) \right) \right. \nonumber \\ && \left. +q^2\left(\frac{M}{p}\right)\left(d_2^a + d^b_2\,\left(\frac{M}{p}\right)^{\frac{1}{2}} 
+ d^c_2\,\left(\frac{M}{p}\right) \right)  
+q\cos\iota\,\left(\frac{M}{p}\right)^{\frac{1}{2}}\left(c_{10}^a + c^b_{10}\,\left(\frac{M}{p}\right) 
+ c^c_{10}\,\left(\frac{M}{p}\right)^{\frac{3}{2}} \right) \right. \nonumber \\ && \left.
+q^2\cos^2\iota\,\left(\frac{M}{p}\right) \left(c_{11}^a
+ c^b_{11}\,\left(\frac{M}{p}\right)^{\frac{1}{2}} + c^c_{11}\,\left(\frac{M}{p}\right) \right)  + \left(\frac{M}{p}\right)^{\frac{5}{2}}\,q^3\,\cos\iota\, \left\{ f^a_7+f^b_7\,\left(\frac{M}{p}\right)^{\frac{1}{2}} \right. \right. \nonumber \\ && \left. \left. + q\,\left(f^a_8+f^b_8\,\left(\frac{M}{p}\right)^{\frac{1}{2}} \right)  + \cos^2\iota\, \left(f^a_9+f^b_9\,\left(\frac{M}{p}\right)^{\frac{1}{2}} \right)  + q\,\cos^2\iota\,\left(f^a_{10}+f^b_{10}\,\left(\frac{M}{p}\right)^{\frac{1}{2}} \right) \right\}
\right] \nonumber \\ \mbox{where } && c_{10}^a = -0.0309341, \qquad c_{10}^b = -22.2416, \qquad c_{10}^c = 7.55265, \qquad c_{11}^a = -3.33476, \qquad c_{11}^b = 22.7013, \nonumber \\ && c_{11}^c = -12.4700, \qquad  f^a_7 = -162.268, \qquad f^b_7 = 247.168, \qquad f^a_8 = 152.125, \qquad f^b_8 = -182.165, \nonumber \\ && f^a_9 = 184.465, \qquad f^b_9 = -267.553, \qquad f^a_{10} = -188.132, \qquad f^b_{10} = 254.067. 
\label{idotfit}
\end{eqnarray}
The pieces of $\dot{L}_{z}$ and $\dot{\iota}$ that are non-vanishing at the pole were constrained to cancel so that the derived $\dot{Q}$ is finite there. The ``d'' coefficients describe these terms. The spin and inclination terms are polynomial in $q$ and $\cos\iota$ and take the form of the next post-Newtonian terms that should enter the flux expressions. The $p$-dependence of the fit starts at 2.5PN order, but the factors of $p$ are not consistent with the post-Newtonian orders of the spin and inclination terms they multiply. The ``c'' coefficients were obtained by a fit to data in the range $p=5M$ to $p=30M$. A fit that was consistent with post-Newtonian expansions was also derived, but this exhibited pathological behaviour when it was used at low $p$. The above form of the fit is virtually identical in the range over which the fit was derived, but does not show the same problems for $p \rightarrow M$. Data was also available for some orbits with $p=3M$. The parts of the fits parameterised by the ``f'' coefficients were derived to reduce the errors in the first fit at $p=3M$ and $p=5M$. The p-dependence was taken to be of a simple form that went to zero asymptotically at least as quickly as the rest of the fit. This ensured the quality of the fit was not degraded at large $p$. Note that several of the ``f'' terms in the $\dot{L_z}$ fit have the same form as the ``c'' coefficient terms. These are separated in \erf{Ldotfit} only because they were derived in these two different ways.

The fit matches the data to an accuracy of $< 3\%
$ at all points. It should be trustworthy in the region $p>3$M, but it will be less accurate as $p \rightarrow$M since data was not used in this range. However, the fit is such that the behaviour in this regime is qualitatively correct, namely $\dot{L}_z < 0$ for near equatorial prograde orbits and $\dot{\iota} > 0$ for all orbits. Inspirals generated using this hybrid scheme will therefore evolve in a reasonable way, but the evolution will not be entirely accurate as the particle gets close to the central black hole. As this last stage of inspiral is very rapid anyway, the loss of accuracy in this regime should not affect the use of the hybrid scheme for exploring the inspiral problem. Moreover, the flux comparisons given in section~\ref{fluxes} indicate that the hybrid scheme does pretty well even for very strong-field orbits. 

As discussed in section~\ref{highinc}, we want to specify $\dot{Q}$ rather than $\dot{\iota}$ in the hybrid scheme. A suitable expression 
for $\dot{Q}$ is obtained by substituting expressions \erf{Ldotfit} and \erf{idotfit} into equation \erf{Qcond}. By including the factor 
$\sin^2\iota/\sqrt{Q}$ in the $\dot{\iota}$ fit, we ensure that the divergent terms cancel precisely and $\dot{Q}$ is 
finite everywhere. The procedure of fitting $\dot{\iota}$ and deriving $\dot{Q}$ may seem more convoluted than fitting 
for $\dot{Q}$ directly, but it ensures that the evolution of the inclination angle is always sensible, and thus generates 
more physically realistic inspirals. We have not presented a fit for $\dot{E}$ here, but this is not necessary since the 
evolution of energy for circular orbits is determined precisely by the evolution of $L_z$ and $Q$ when we apply the circular 
orbit correction described in section~\ref{nearcirc}.

This procedure has allowed us to improve the hybrid fluxes for circular orbits, but we have not yet improved the eccentricity-dependent terms in the fluxes. However, as discussed earlier, we must include the usual $(1-e^2)^{3/2}$ prefactor to ensure reasonable behaviour in the limit 
$e\rightarrow 1$. The resulting expression for $\dot{L}_z$ is
\begin{equation}
\left(\dot{L}_z\right)_{\rm mod} = (1-e^2)^{\frac{3}{2}} \left[(1-e^2)^{-\frac{3}{2}}\left(\dot{L}_z \right)_{\rm 2PN} (p,\iota,e,a) 
- \left(\dot{L}_z \right)_{\rm 2PN} (p,\iota,0,a) + \left(\dot{L}_z 
\right)_{\rm fit} \right]
\label{Ldotfinal}
\end{equation}
in which the subscript ``2PN'' refers to expression \erf{new_Ldot_2} and the subscript ``fit'' refers to expression \erf{Ldotfit}. We obtain the expression for $\dot{Q}/\sqrt{Q}$ in the same way
\begin{eqnarray}
\left(\dot{Q}\right)_{\rm mod} &=& (1-e^2)^{\frac{3}{2}} \sqrt{Q(p,\iota,e,a)} \left[(1-e^2)^{-\frac{3}{2}}\left(\frac{\dot{Q}}{\sqrt{Q}} \right)_{\rm 2PN} (p,\iota,e,a) - \left(\frac{\dot{Q}}{\sqrt{Q}} \right)_{\rm 2PN} (p,\iota,0,a) \right. \nonumber \\ &&\left. + 2\,\tan\iota\left(\left(\dot{L}_z \right)_{\rm fit} 
+\frac{\sqrt{Q(p,\iota,0,a)}}{\sin^2\iota}\left(\dot{\iota}\right)_{\rm fit}\right)\right] .
\label{Qdotfinal}
\end{eqnarray}
The final expression for $\dot{E}$ is still obtained from equation \erf{Edotcorr} and expression \erf{new_Edot_2}.

%%%%%%%%%%%%%%%%%%%%%%%%%%%%%%%%%%%%%%%%%%%%%%%%%%%%%%%%%%%%%%%%%%%%%%%%%%%%%%%%%5

\subsubsection{Eccentric orbits}
\label{eccterms}
The expressions quoted above are missing some eccentricity-dependent pieces. The omission of these eccentricity terms manifests itself as some unphysical behaviour for inspirals generated using the hybrid scheme. For prograde equatorial orbits with radius $p=2$M around a black hole of spin $a=0.99$M, we find that the magnitude of $\dot{L}_{z}$ becomes smaller as the orbital eccentricity increases, in stark contrast to the behaviour of Teukolsky generated fluxes (see Table~\ref{fluxcomp}). For eccentricity $e>0.3$ we find $\dot{L}_{z} > 0$ and $\dot{p} >0$ which is unphysical. Glampedakis and Kennefick \cite{kgdk} tabulate Teukolsky-based fluxes for equatorial eccentric orbits and these can be used to derive an additional eccentricity-dependent correction to the fluxes. However, the inclination dependence of this correction is unknown since the data is provided for equatorial orbits only, and it is difficult to constrain the $p$ dependence with the limited data provided in \cite{kgdk}. Recently, Sago \etal published new 2PN results for the energy and angular momentum fluxes from orbits with small eccentricity {\em and} small inclination \cite{sago05b}. After rewriting the expressions in our coordinates, these results confirm the extrapolations that lead us to equations \erf{new_Edot_2}-\erf{new_Ldot_2} and also give the eccentricity dependence of one of the 2PN terms that we could not otherwise derive. We deduce that equations \erf{new_Edot_2}-\erf{new_Ldot_2} should only change in the final term, with the modifications
\begin{eqnarray}
-\frac{527}{96} q^2 \left( \frac{M}{p}\right)^2 \sin^2\iota \qquad &\rightarrow& \qquad -\left(\frac{527}{96} + \frac{6533}{192} e^2\right) q^2 \left( \frac{M}{p}\right)^2 \sin^2\iota \qquad \mbox{in equation \erf{new_Edot_2}} \nonumber \\
-\frac{45}{8} q^2 \left( \frac{M}{p}\right)^2 \cos\iota \sin^2\iota \qquad &\rightarrow& \qquad -\left(\frac{45}{8} + \frac{37}{2} e^2\right) q^2 \left( \frac{M}{p}\right)^2 \cos\iota\sin^2\iota \qquad \mbox{in equation \erf{new_Ldot_2}} \nonumber
\end{eqnarray}
When these expressions (adding the consistency corrections and the fit for the circular flux) were tested by comparison to Teukolsky-based results, they were found to perform less well than the version of the hybrid scheme described above. It is not clear why this is the case. The new results in \cite{sago05b} are inconsistent at 2.5PN order with the results in \cite{chapter}. Although we do not use the 2.5PN order results, it is possible the inconsistency is indicative of some other error which might also have affected the new eccentricity terms above. However, it may just be that the new expressions do not perform well because we are using a weak-field expansion (the PN expansion) and applying it in the strong-field. The motivation for developing the hybrid scheme was to find a set of expressions that reproduce accurate, Teukolsky-based results as closely as possible. In that spirit, we recommend ignoring these new corrections and using the expressions quoted in sections~\ref{2PN}-\ref{Teukfit} instead. Once the new results have been more thoroughly tested, and perhaps augmented with further fits to Teukolsky-based data, it might be possible to improve our expressions further, but this will be considered in future publications.

Recent work by Sago \etal \cite{sago05}, based on a paper by Mino \cite{mino03}, provides an expression for the time evolution of the Carter constant, $Q$, which uses the same Teukolsky variables needed for the computation of the energy and angular momentum fluxes. The paper \cite{sago05b} makes use of this expression to derive a 2PN formula for $\dot{Q}$ as well. The analogue of equation~\erf{Qdot_new} is found to be
\begin{eqnarray}
\dot{Q} &=& -\frac{64}{5} \frac{\mu^2}{M} \left(\frac{M}{p}\right)^{7/2} \,\sqrt{Q}\,\sin\iota\,
(1-e^2)^{3/2}\,\left[ g_9(e) - q\left(\frac{M}{p}\right)^{3/2} \cos\iota g_{10}^{b}(e) -\left(\frac{M}{p}\right) g_{11}(e)
\right. \nonumber \\ && \left. + \pi\left(\frac{M}{p}\right)^{3/2} g_{12}(e) 
-\left(\frac{M}{p}\right)^2 g_{13}(e)
+ q^2\left(\frac{M}{p}\right)^2 \,\left(\frac{211}{96} + \frac{847}{96}\,e^2\right)  \right]. \label{Qdot_sago}
\end{eqnarray}
Note that Sago \etal actually give a 2PN expression for $\dot{Q}$, not $\dot{Q}/\sqrt{Q}$ as given above. We quote the above result, obtained by expanding $\sqrt{Q}$ at 2PN order and comparing to Sago \etals results, for consistency with \erf{Qdot_new}. Also note that Sago \etals results were computed to linear order in $\iota^2$, so it is not possible to determine any pieces proportional to $\sin^2\iota$ in the final term of equation \erf{Qdot_sago}, although such terms are present in expression \erf{Qdot_new}. 

Equation~\erf{Qdot_sago} shows amazing agreement with our previous result, considering the way in which the latter was derived and this gives us some confidence in our approach. At present, there is no strong-field Teukolsky-based data for $\dot{Q}$, so we have no way of determining if equation \erf{Qdot_sago} performs better than \erf{Qdot_new}. Due to the inconsistencies in \cite{sago05b} mentioned above, at present we suggest ignoring \erf{Qdot_sago} and using \erf{Qdot_new} (more precisely, \erf{Qdot_new} augmented with the circular fit \erf{Qdotfinal}). As Teukolsky data based on the new formula for $\dot{Q}$ \cite{sago05} becomes available, we will be able to assess the alternative expressions for $\dot{Q}$ and determine which gives the best results in the hybrid scheme.

The results given in the following section are based on equations \erf{new_Edot_2}-\erf{new_Ldot_2} and \erf{Qdot_new}, augmented with the near-circular correction \erf{circcond} and the fits to circular Teukolsky data \erf{Ldotfit}-\erf{Qdotfinal}. We recommend using this form of the hybrid scheme in calculations, although one should bear in mind that the eccentricity pieces in these fluxes are incomplete and the fluxes are unphysical in a few small regions of parameter space. However, the unphysical behaviour appears only very close to plunge, thus the hybrid scheme in its current form should be adequate for most applications. Once Teukolsky-based data is available for a wider range of generic orbits it will be possible to compute a significantly more accurate fit to the missing eccentricity-dependent pieces. As we saw earlier, it is likely that fitting only the $e^2$ pieces of Teukolsky-based fluxes will be sufficient to give accurate inspirals and this will be pursued in the future.

%%%%%%%%%%%%%%%%%%%%%%%%%%%%%%%%%%%%%%%%%%%%%%%%%%%%%%%%%%%%%%%%%%%%%%%%%%%%%%%%%%%%%%%%%%%%%%%%%%%%%%%%%%%%%%%%%%%%%%

\section{Inspiral properties}
\label{applications}

\subsection{Stability of nearly circular orbits}
\begin{figure}
\centerline{\includegraphics[keepaspectratio=true,height=5in,angle=-90]{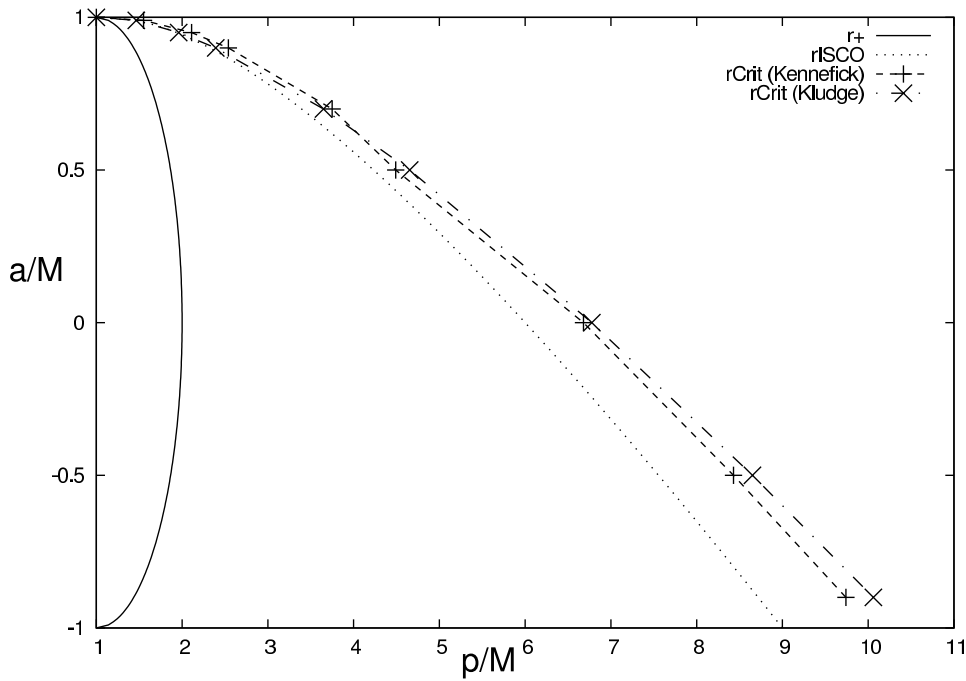}}
\caption{Location of the critical radius $r_{crit}$, at which the $\dot{e} =0$ curve intersects the $e=0$ axis for equatorial orbits, as a function of spin. The pluses mark points computed using Teukolsky fluxes by Kennefick \cite{kenn98}, while the crosses indicate points computed using the hybrid scheme described in this paper. The agreement between the two curves is within $\sim5\%$ for all spins.}
\label{circstab}
\end{figure}

Kennefick \cite{kenn98} examined the stability of nearly circular orbits in the equatorial plane using Teukolsky-based calculations. This involved expanding the eccentricity derivative $\dot{e}$ near $e=0$. As described above, we expect $\dot{e} \approx f(p)\,e$ near $e=0$. For orbits very close to the separatrix, $\dot{e} > 0$ as $e \rightarrow 0$, while for orbits with larger periapse, $\dot{e} < 0$ in that limit. It is possible to find the point where the transition in behaviour occurs, i.e., where $f(p) =0$. The locus of points where $\dot{e} =0$ is important for determining the global properties of inspiral trajectories, and the root $f(p)=0$ is the point where this curve intersects the $e=0$ axis. Figure~\ref{fig4} indicates that the improvements to the hybrid scheme move the point where $\dot{e}$ changes sign in a given inspiral much closer to plunge when compared to the original GHK scheme. This is in keeping with the results presented in \cite{kenn98}. We can compute the root $f(p)=0$ as a function of the black hole spin, $a/M$, for the new hybrid fluxes and compare this to Kennefick's results (these were corrected in \cite{kgdk} and we use the corrected values in this plot). This comparison is shown in Figure~\ref{circstab}. We find that the agreement in the location of the critical radius is extremely good, within $\sim5\%
$ for all values of the black hole spin. This is particularly remarkable since the location of the critical radius depends on the leading eccentricity terms in the fluxes, and we have already seen that the 2PN expressions used are somewhat inaccurate close to the separatrix. The structure of the inspiral phase space depends significantly on two curves -- the separatrix at which orbits plunge, and the critical curve $\dot{e}=0$, where inspirals turn up. The hybrid scheme includes the correct separatrix by construction and Figure~\ref{circstab} indicates that the new hybrid fluxes recover the critical curve very well, so we can have some confidence that inspirals generated under this improved hybrid scheme are qualitatively correct.

%%%%%%%%%%%%%%%%%%%%%%%%%%%%%%%%%%%%%%%%%%%%%%%%%%%%%%%%%%

\subsection{Flux comparisons}
\label{fluxes}
The hybrid flux expressions allow us to compute the rates of loss of energy and angular momentum from an arbitrary geodesic orbit. The true fluxes are given by solution of the Teukolsky equation, and we can compare the hybrid results to existing Teukolsky-based results available in the literature. This comparison is shown in Table~\ref{fluxcomp}. The table is divided into three sections. The top section is for circular and inclined orbits, for which the Teukolsky data was provided by Scott Hughes \cite{scott}. The middle section is for eccentric and equatorial orbits, for which the Teukolsky data was taken from Glampedakis and Kennefick \cite{kgdk}. The bottom section is for generic orbits, and the Teukolsky data was taken from Drasco and Hughes \cite{drasco05}. In each case we tabulate the hybrid and Teukolsky-based fluxes of energy and angular momentum, plus the rate of change of inclination angle. Drasco and Hughes use the ``spherical'' formula \erf{GHKidot} to evolve the inclination angle, which we have seen to be inaccurate. The Teukolsky results should not therefore be regarded as more accurate for that comparison.

Overall, the agreement between the hybrid and Teukolsky results is remarkably good. The agreement for circular, inclined orbits is excellent, but this is unsurprising since we used Teukolsky data to fit the circular part of the hybrid fluxes. The hybrid scheme also performs very well for orbits of low eccentricity, but the performance is a little worse for orbits of higher eccentricity, particularly in the strong-field. Specifically, we see the decrease in $|\dot{L}_{z}|$ with increasing eccentricity for $p=2$M and $p=3$M that was discussed earlier, and find one prograde equatorial orbit that has $\dot{L}_{z}>0$. This is a consequence of the missing eccentricity-dependent terms in the fluxes. In the future, deriving a fit to the missing terms using Teukolsky-based data should correct these remaining problems.

\begin{table}
\begin{tabular}{c|c|c|c|c|c|c|c|c|c}
\hline $a/M$&$p/M$&$e$&$\iota$&\multicolumn{2}{|c} {$\left(M/\mu\right)^2\,\dot{E}$}&\multicolumn{2}{|c} {$\left(M/\mu^2\right)\, \dot{L}_{z}$} &\multicolumn{2}{|c}
{$\left(M^3/\mu^2\right)\,\dot{\iota}$}\\ 
\hline
& & & & Teuk & Hybrid & Teuk & Hybrid & Teuk & Hybrid \\ \hline $0.05$&$100$&$0$&$60$&$-6.237\times10^{-10}$&$-6.237\times10^{-10}$&$-3.119\times10^{-7}$&$-3.119\times10^{-7}$&$6.706\times10^{-12}$&$6.718\times10^{-12}$\\ $0.95$&$100$&$0$&$60$&$-6.219\times10^{-10}$&$-6.219\times10^{-10}$&$-3.122\times10^{-7}$&$-3.122\times10^{-7}$&$1.267\times10^{-10}$&$1.268\times10^{-10}$\\ $0.05$&$7$&$0$&$60$&$-3.951\times10^{-4}$&$-3.958\times10^{-4}$&$-3.697\times10^{-3}$&$-3.704\times10^{-3}$&$1.104\times10^{-5}$&$1.107\times10^{-5}$\\ $0.95$&$7$&$0$&$60$&$-3.055\times10^{-4}$&$-3.051\times10^{-4}$&$-3.368\times10^{-3}$&$-3.362\times10^{-3}$&$1.706\times10^{-4}$&$1.701\times10^{-4}$\\ \hline
$0.5$&$5$&$0.1$&$0$&$-1.813\times10^{-3}$&$-1.787\times10^{-3}$&$-2.063\times10^{-2}$&$-2.019\times10^{-2}$&$0$&$0$\\ $0.5$&$5$&$0.2$&$0$&$-2.087\times10^{-3}$&$-1.951\times10^{-3}$&$-2.208\times10^{-2}$&$-2.019\times10^{-2}$&$0$&$0$\\
$0.5$&$5$&$0.3$&$0$&$-2.601\times10^{-3}$&$-2.170\times10^{-3}$&$-2.480\times10^{-2}$&$-2.001\times10^{-2}$&$0$&$0$\\ $0.99$&$2$&$0.1$&$0$&$-4.4067\times10^{-2}$&$-3.938\times10^{-2}$&$-1.657\times10^{-1}$&$-1.204\times10^{-1}$&$0$&$0$ \\ $0.99$&$2$&$0.2$&$0$&$-4.723\times10^{-2}$&$-4.702\times10^{-2}$&$-1.700\times10^{-1}$&$-7.249\times10^{-2}$&$0$&$0$ \\ $0.99$&$2$&$0.3$&$0$&$-5.444\times10^{-2}$&$-5.084\times10^{-2}$&$-1.771\times10^{-1}$&$2.395\times10^{-3}$&$0$&$0$ \\
$0.99$&$3$&$0.1$&$0$&$-1.083\times10^{-2}$&$-1.096\times10^{-2}$&$-6.583\times10^{-2}$&$-6.293\times10^{-2}$&$0$&$0$ \\ $0.99$&$3$&$0.2$&$0$&$-1.153\times10^{-2}$&$-1.259\times10^{-2}$&$-6.684\times10^{-2}$&$-6.001\times10^{-2}$&$0$&$0$ \\ $0.99$&$3$&$0.3$&$0$&$-1.262\times10^{-2}$&$-1.439\times10^{-2}$&$-6.825\times10^{-2}$&$-5.473\times10^{-2}$&$0$&$0$ \\ $0.99$&$11$&$0.1$&$180$&$-4.961\times10^{-5}$&$-4.932\times10^{-5}$&$1.736\times10^{-3}$&$1.711\times10^{-3}$&$0$&$0$ \\ $0.99$&$11$&$0.2$&$180$&$-5.589\times10^{-5}$&$-5.248\times10^{-5}$&$1.821\times10^{-3}$&$1.709\times10^{-3}$&$0$&$0$ \\ $0.99$&$11$&$0.3$&$180$&$-6.657\times10^{-5}$&$-5.687\times10^{-5}$&$1.963\times10^{-3}$&$1.695\times10^{-3}$&$0$&$0$ \\ \hline
$0.9$&$6$&$0.1$&$40.192285$&$-6.145\times10^{-4}$&$-6.196\times10^{-4}$&$-7.551\times10^{-3}$&$-7.534\times10^{-3}$&$0$&$2.611\times10^{-4}$ \\ $0.9$&$6$&$0.3$&$40.176668$&$-7.209\times10^{-4}$&$-7.512\times10^{-4}$&$-7.641\times10^{-3}$&$-7.632\times10^{-3}$&$0$&$3.197\times10^{-4}$ \\ $0.9$&$6$&$0.5$&$40.145475$&$-8.654\times10^{-4}$&$-8.475\times10^{-4}$&$-7.670\times10^{-3}$&$-7.143\times10^{-3}$&$0$&$3.768\times10^{-4}$ \\
$0.9$&$6$&$0.1$&$80.046323$&$-8.060\times10^{-4}$&$-8.007\times10^{-4}$&$-3.427\times10^{-3}$&$-3.395\times10^{-3}$&$0$&$4.371\times10^{-4}$ \\ 
$0.9$&$6$&$0.3$&$80.042690$&$-1.086\times10^{-3}$&$-9.879\times10^{-4}$&$-4.023\times10^{-3}$&$-3.659\times10^{-3}$&$0$&$5.051\times10^{-4}$ \\
$0.9$&$6$&$0.5$&$80.035363$&$-1.685\times10^{-3}$&$-1.163\times10^{-3}$&$-5.133\times10^{-3}$&$-3.761\times10^{-3}$&$0$&$5.602\times10^{-4}$ \\ \hline
\end{tabular}
\caption{Comparison of hybrid and Teukolsky-based results. We tabulate the fluxes of energy and angular momentum and the rate of change of the inclination angle, $\iota$, for a variety of geodesics. These are divided into circular/inclined (top section), eccentric/equatorial (middle section) and eccentric/inclined (bottom section).}
\label{fluxcomp}
\end{table}

%%%%%%%%%%%%%%%%%%%%%%%%%%%%%%%%%%%%%%%%%%%%%%%%%%%%%%%%%%

\subsection{Sample inspirals}
The main application of the hybrid scheme is to generate inspiral trajectories. These can be used to investigate the general properties of extreme mass ratio inspirals, and also as input for the generation of approximate gravitational waveforms \cite{kludge_paper}. Such approximate waveforms are being used for scoping out data analysis issues for the detection of EMRIs with LISA \cite{gair04} and may also find application as detection templates in the final analysis of LISA data.

\begin{figure}
\centerline{\includegraphics[keepaspectratio=true,height=5in,angle=-90]{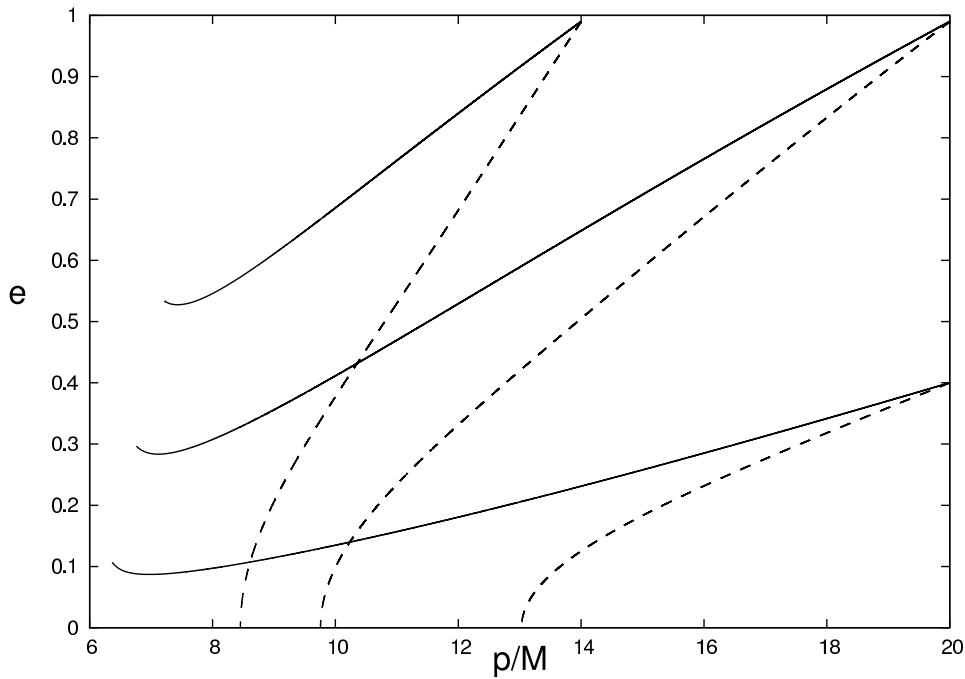}}
\caption{Inspirals in the ($p$, $e$) plane computed using the revised hybrid scheme (solid curves). The dashed curves represent the original GHK results and are included for comparison. Initial values for $(p,e,\iota)$ are $(20M,0.99,100^\circ), (14M,0.99,100^\circ)$ and $(20M,0.4, 100^\circ)$. The black hole spin is $a=0.5M$.}
\label{FinalInspComp}
\end{figure}

\begin{figure}
\centerline{\includegraphics[keepaspectratio=true,height=5in,angle=-90]{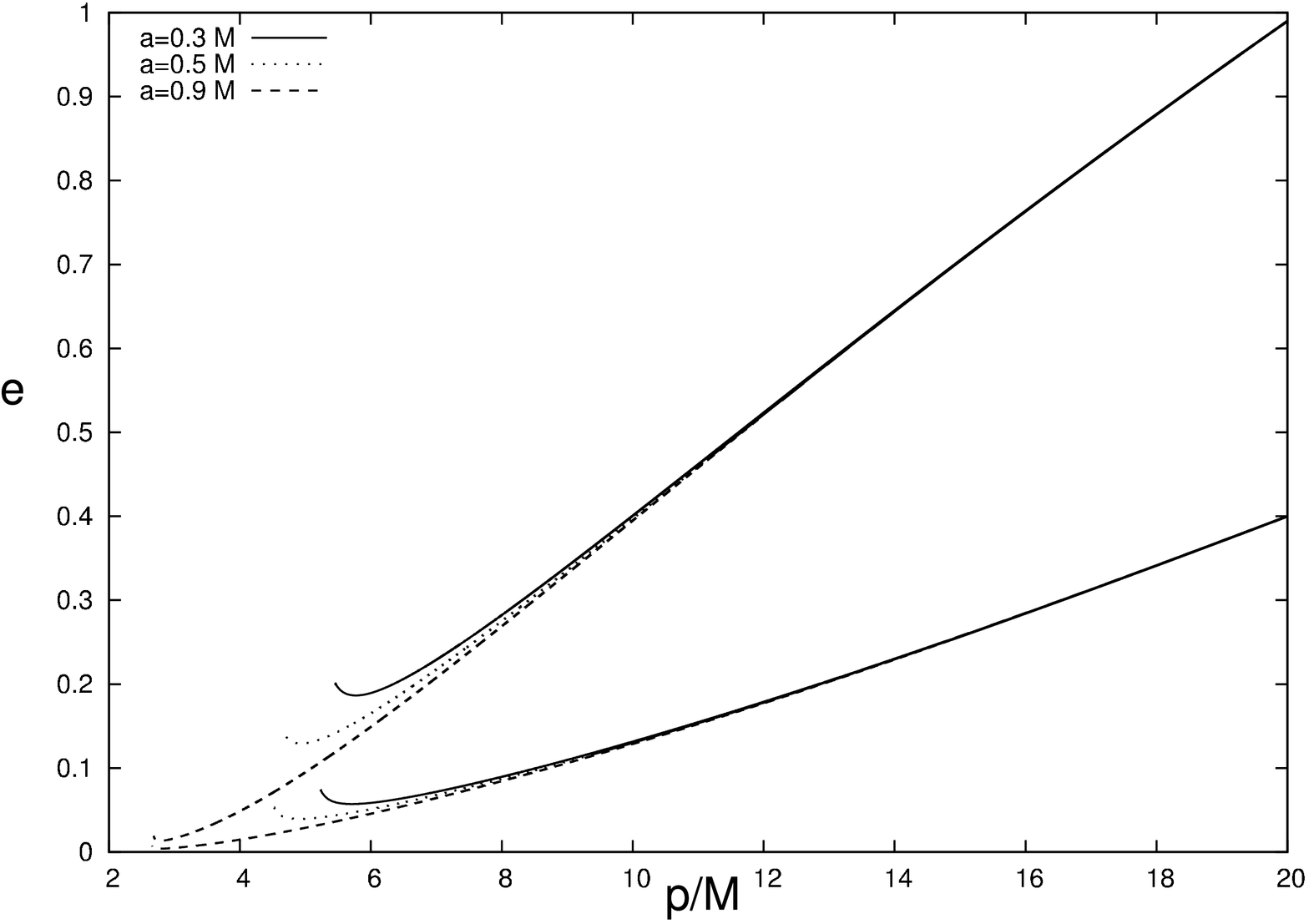}}
\centerline{\includegraphics[keepaspectratio=true,height=5in,angle=-90]{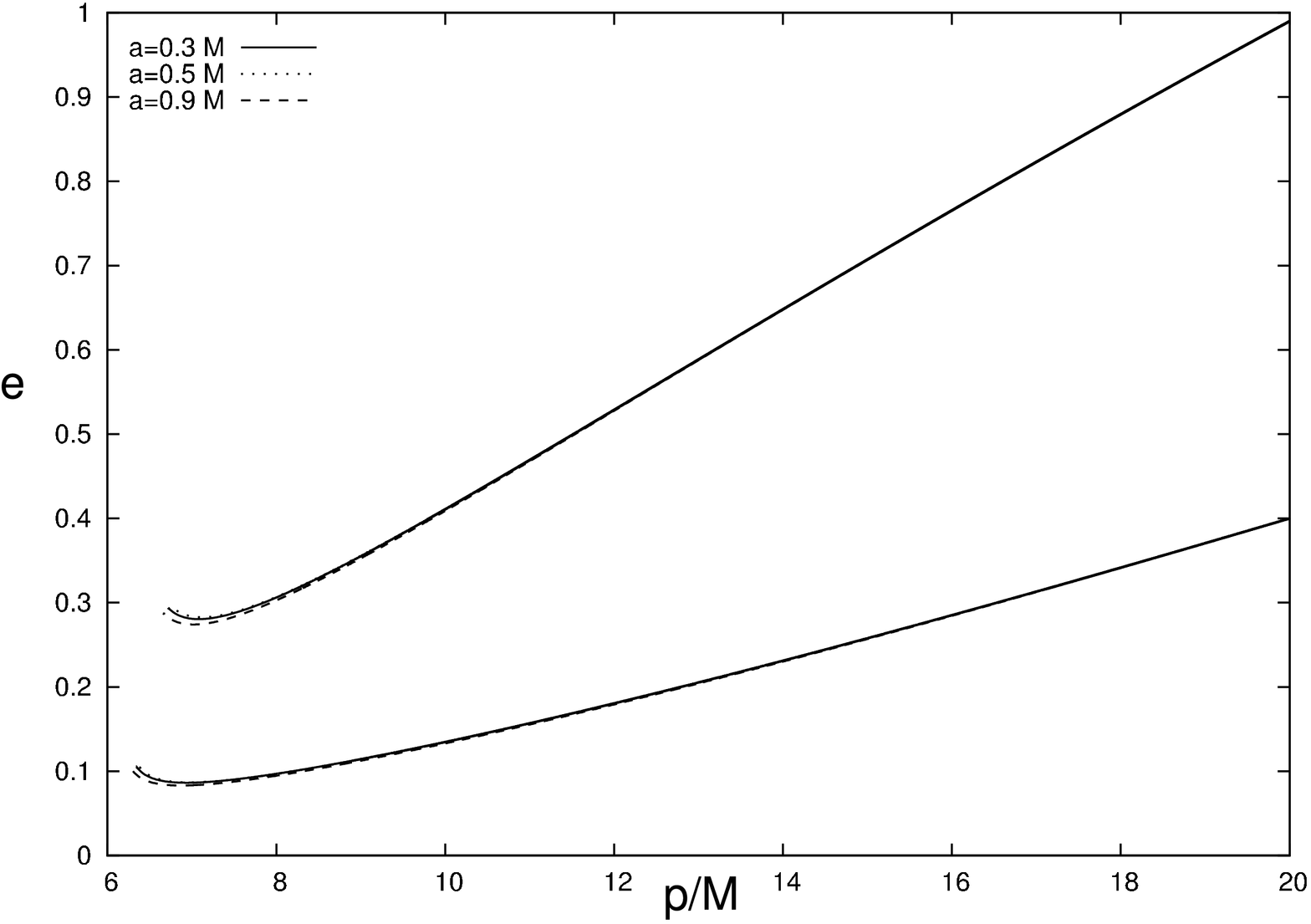}}
\caption{A set of inspirals in the $p-e$ plane computed using the revised hybrid scheme. For each inspiral we initially set $p= 20$M and $\iota = 30^\circ$ (top panel) or $\iota= 100^\circ$ (bottom panel). The inspirals are shown for two different initial eccentricities, $e=0.4$ or $e=0.99$, and three different values of the central black hole spin, $a=0.3$M (solid curves), $a=0.5$M (dotted curves) and $a=0.9$M (dashed curves).}
\label{peInsp}
\end{figure}

In Figure~\ref{FinalInspComp} we illustrate some inspirals in the $p-e$ plane computed using the final form of the hybrid scheme \erf{Ldotfinal} for a black hole of spin $a=0.5$M. We show the same inspirals as computed under the original GHK scheme for comparison. The improvement of the new scheme over the old one is quite evident and the old pathologies are no longer present. In Figure~\ref{peInsp} we show a further sequence of inspirals in the $p-e$ plane computed under the new hybrid scheme. These all have initial semi-latus rectum of $p=20$M, and have a variety of eccentricities, inclinations and spins. Every inspiral has the same basic structure $\--$ initially the orbit circularises as it inspirals, but once the orbit gets close to plunge, a critical point is reached where $\dot{e}=0$ and the eccentricity then begins to increase until the object plunges into the black hole. Orbits of lower initial eccentricity circularise less rapidly. As the spin of the central black hole is increased, the plunge point moves closer to the central black hole, and so does the critical radius. Orbits of higher inclination plunge further from the central black hole, so the net effect of spin is reduced, as is clear from the $\iota =100^{\circ}$ panel in Figure~\ref{peInsp}.

\begin{figure}
\centerline{\includegraphics[keepaspectratio=true,height=5in,angle=-90]{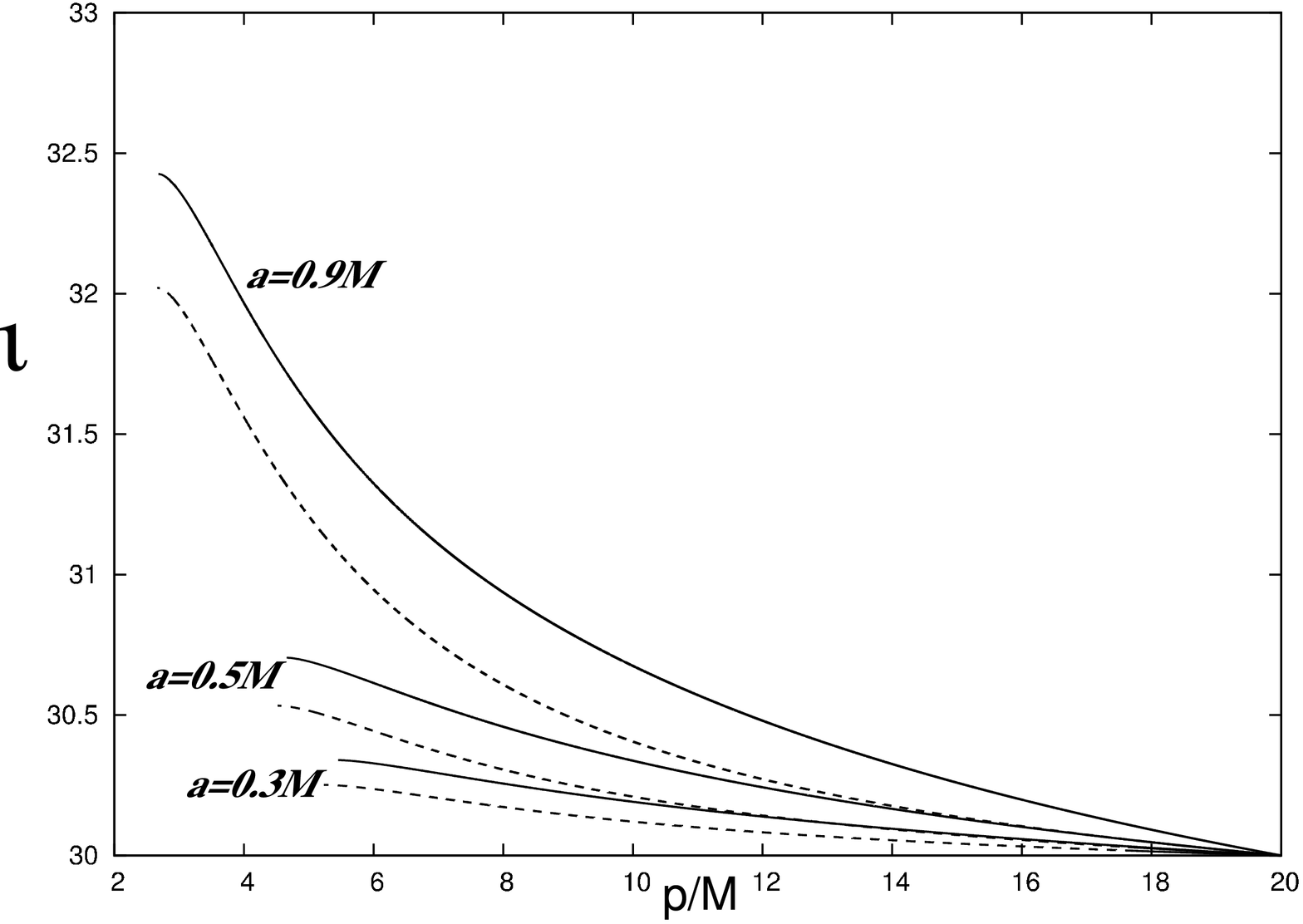}}
\centerline{\includegraphics[keepaspectratio=true,height=5in,angle=-90]{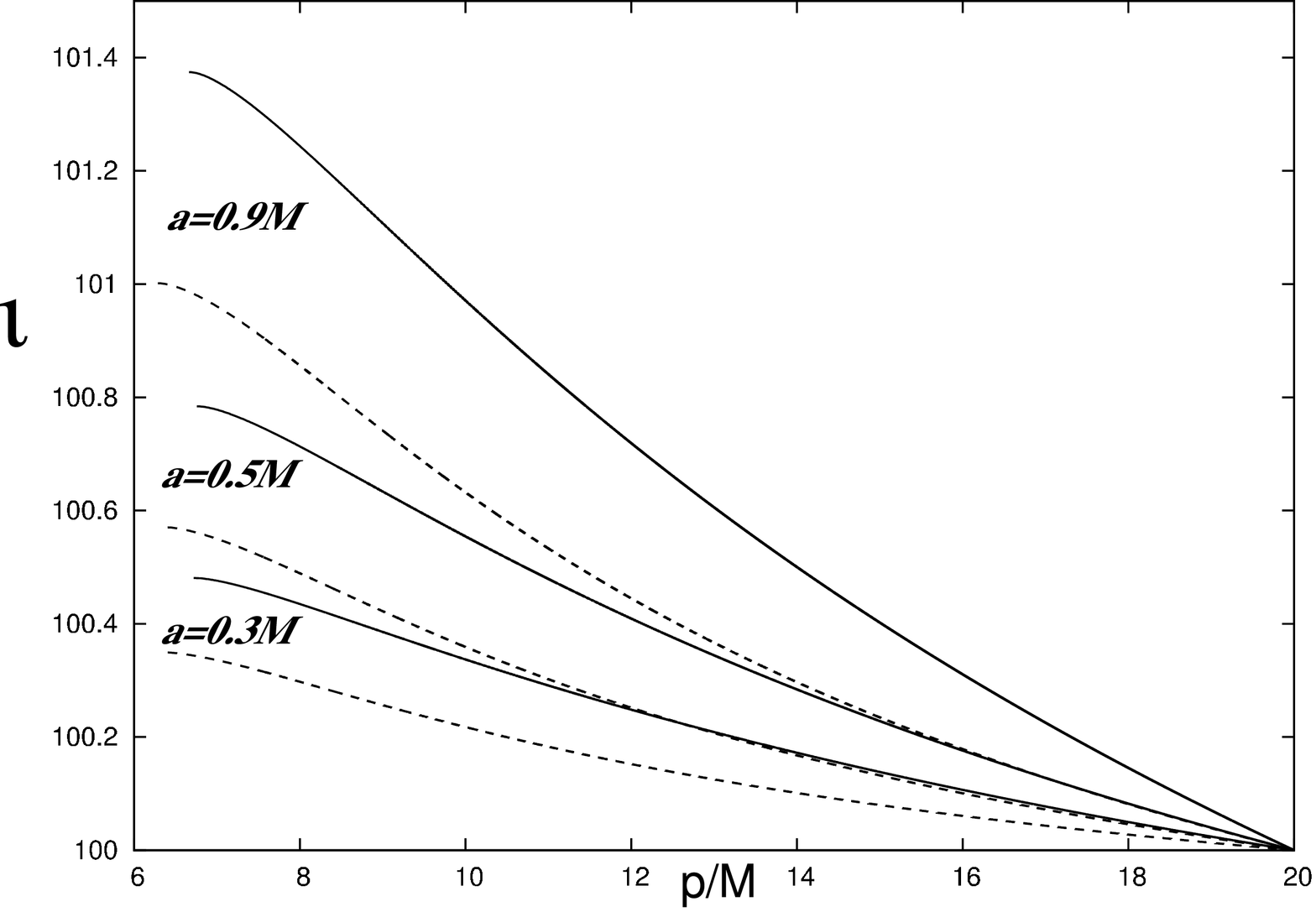}}
\caption{The radiation reaction induced evolution of $\iota$ during inspiral for the inspirals shown in Figure~\ref{peInsp}. For each inspiral we initially set $p= 20$M and $\iota = 30^\circ$ (top panel) or $\iota= 100^\circ$ (bottom panel). We used two different values of the initial eccentricity, $e=0.4$ (dashed curves) or $e=0.99$ (solid curves), and three values for the black hole spin (as labelled).}
\label{piInsp}
\end{figure}

Figure~\ref{piInsp} illustrates the same inspirals, but shows how the inspiral proceeds in the $p-\iota$ plane. We see that, typically, $\iota$ changes only slightly over the inspiral, but always increases. The magnitude of this change increases as the spin is increased, and is greater for orbits nearer to polar ($\iota=90^{\circ}$) than for nearly equatorial ($\iota=0$) orbits. Orbits of lower initial eccentricity also exhibit a lower total change in inclination over the inspiral. For low spins, the change in $\iota$ is less than $1^\circ$, but it can be somewhat larger ($\sim 3^\circ$) for near polar inspirals into rapidly rotating black holes. These features, and the non-zero, but small, increase in inclination is in good agreement with the results of Teukolsky-based calculations \cite{scott2}. 

%%%%%%%%%%%%%%%%%%%%%%%%%%%%%%%%%%%%%%%%%%%%%%%%%%%%%%%%%

\section{Conservative self-force corrections}
\label{conscorr}
In this paper we have focused on improving the computation of the trajectories that EMRIs follow in phase space, specifically the evolution of the three principal constants of the motion, $E$, $L_z$ and $Q$. These three constants determine the shape of any given geodesic. However, a geodesic also has three additional ``positional'' constants of the motion, which essentially label the position of the particle along the geodesic trajectory at some fiducial time. The self-force interaction between the inspiralling body and the central black hole acts to change not only the principal constants of the motion (the dissipative self-force), but also the positional constants (the conservative self-force). In this analysis we have looked only at the dissipative contribution. However, the influence of the conservative self-force on the orbital evolution cannot be ignored, since it could lead to several cycles of phase discrepancy in template waveforms over the course of an inspiral. In a recent paper \cite{pound05} Pound \etal explored a toy problem in which an orbiting charged particle experienced both a dissipative and a conservative electromagnetic self-force. They found that in the weak-field the phasing contribution from the conservative self-force was comparable to or larger than the dissipative contribution. While that particular analysis was not applicable in the strong-field regime and did not look at the gravitational self-force directly, it is likely that the conservative gravitational self-force will have a significant influence on the phasing of an inspiral.

The conservative self-force alters the orbit that an inspiralling body with particular energy etc. follows. This change in the orbit in principle will modify the radiated fluxes of $E$, $L_z$ and $Q$. However, this is a much lower order effect and so can be ignored. The expressions presented here can therefore be accurately used to model EMRI phase space trajectories. On the other hand, conservative effects should be included when an inspiral path (i.e., the path through space which the orbiting body follows) is computed based on the phase space evolution. This is the next step in constructing approximate waveforms (see discussion in next section and Figure~\ref{snapshots}). One way to add conservative effects is to include the three positional constants of the motion when parameterising the orbit and evolve these as well as the principal constants along the phase space trajectory. This is discussed in more detail in \cite{kludge_paper}. Perturbative calculations have not yet reached the stage of computing the conservative self-force for strong-field orbits, although this should be possible in the near future for Schwarzschild orbits \cite{poisson}. There are post-Newtonian expressions for the conservative self-force available in the literature (e.g. \cite{junker92}), so it should be possible to put together existing results to construct a hybrid scheme for the evolution of the positional orbital constants, along similar lines to the hybrid scheme described in this paper for the principal constants. However, this is a complicated procedure and we leave it for a future paper.

%%%%%%%%%%%%%%%%%%%%%%%%%%%%%%%%%%%%%%%%%%%%%%%%%%%%%%%%%

\section{Concluding remarks}
\label{conc}
In this paper we have described a number of ways to alter the GHK hybrid scheme in order to expand its validity and reliability. The inclusion of all of these corrections leads to inspirals that are physically reasonable and seem qualitatively correct throughout parameter space. Moreover, direct comparisons of the hybrid fluxes to Teukolsky-based results indicate that the approach performs well. However, at present we can only judge the scheme based on physical intuition and lessons learnt from studies of specific families of orbits (equatorial-eccentric and circular-inclined) and the limited number of fluxes for generic (inclined-eccentric) orbits that are currently available \cite{drasco05}. As more extensive results for generic orbits are generated in the near future, we will be able to more thoroughly test this new hybrid scheme and improve it if necessary. An important point to note is that the new scheme is still computationally inexpensive, which is why it is valuable to pursue this approach in conjunction with more accurate perturbative calculations.

There are a number of ways in which the scheme might be further improved in the future. The $\dot{E}$, $\dot{L}_{z}$ and $\dot{Q}$ flux expressions, 
\erf{new_Edot_2}--\erf{new_Ldot_2} and \erf{Qdot_new}, could be improved by including additional terms. The current expressions do not include all eccentricity terms, and we saw in section~\ref{eccterms} that additional terms currently available in the literature can be easily included, although these do not appear to improve the performance of the hybrid scheme. However, it should hopefully still be possible to gain improvements by using higher order PN results as these become available. The spin-independent terms could also be amended by using the 2PN fluxes derived by Gopakumar and Iyer \cite{gopu} for a binary system with non-spinning members moving in quasi-elliptical orbits. A modification like the latter should improve the numerical inspirals for $ e \sim 1 $. Another approach would be to construct the Pad\'e approximants of the PN fluxes, which are typically more effective than the original Taylor-series fluxes \cite{pade}. Once generic Teukolsky data is available, the scheme could also be enhanced by improving the eccentricity-dependent pieces of the
$\dot{E}$, $\dot{L}_{z}$ and $\dot{\iota}$ fluxes using fits to the Teukolsky results. As discussed in section~\ref{conscorr}, a significant gap in the present hybrid scheme is the omission of conservative self-force corrections. While this omission does not affect the phase space inspiral trajectories described here, it might significantly influence the gravitational waveform phasing and can not therefore be ignored. Using PN results and data from perturbative self-force calculations, it should be possible to include the conservative self-force by computing a ``positional'' phase space trajectory in conjunction with the ``principal'' phase space trajectory described here. This will be pursued in the future.

A key point to remember in any future version of the hybrid scheme is that the consistency condition \erf{circcond} must be satisfied by the fluxes in order to produce reasonable inspirals. To incorporate this correction, we added a term to the energy flux that is essentially the difference between the prescribed energy flux and what it should be to be consistent with the prescribed angular momentum and inclination fluxes. As the flux expressions are pushed to higher and higher order, the size of this correction will naturally diminish and we will ultimately converge to the true adiabatic inspiral.

The conditions \erf{circcond} and \erf{polarcond} also have some relevance for Teukolsky-based calculations for generic orbits. Until recently, these had been carried out only for circular orbits of arbitrary inclination \cite{scott} and equatorial orbits of arbitrary eccentricity \cite{kgdk}. In both cases, the additional symmetries allow the Carter constant to be evolved correctly. More recently, Teukolsky-based calculations have been carried out for ``snapshots'' of generic orbits \cite{drasco05}. Due to the difficulties in computing the evolution of the Carter constant for a generic inspiral, these results made use of the constant inclination approximation, $\dot{\iota}=0$, to compute $\dot{Q}$. It is clear from the analysis here that this will not work for nearly circular or nearly polar orbits, since we need the correct $\dot{\iota}$ in those limits to ensure conditions \erf{circcond} and \erf{polarcond} are satisfied. The constant inclination approach is therefore problematic. However, the new formula for the evolution of $Q$ presented by Sago {\em et al.} \cite{sago05} allows computation of $\dot{Q}$ using the same Teukolsky variables used to evaluate $\dot{E}$ and $\dot{L}_z$. Future generic codes should make use of this new formula to avoid the consistency problems described here.

\begin{figure}
\centerline{\includegraphics[keepaspectratio=true,height=7in,angle=-90]{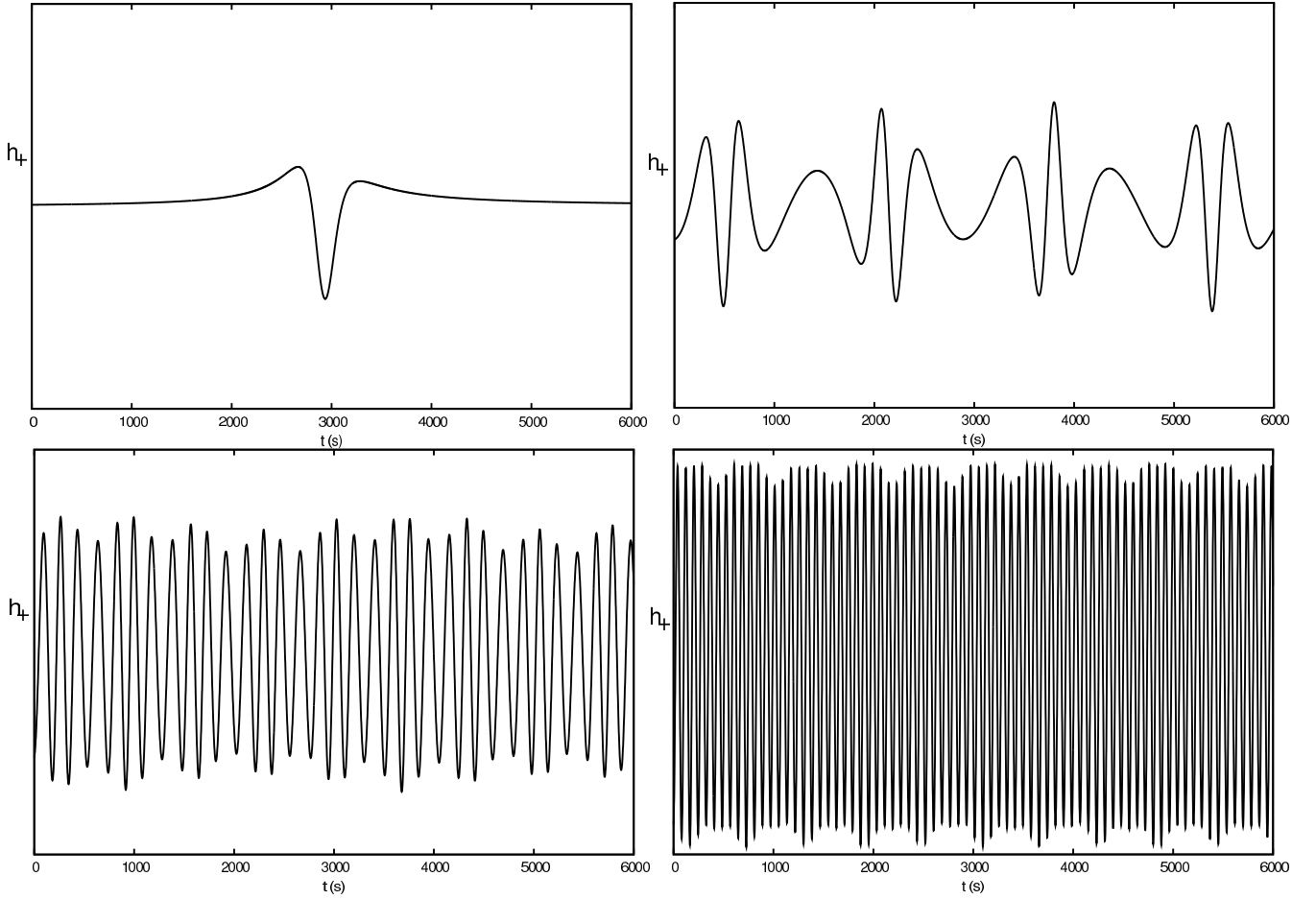}}
\caption{Snapshots of an approximate gravitational waveform from an 
inspiral with $a=0.9$M, $M=10^6$\msun, $\mu=1$\msun \,and initial orbital parameters $p=20$M, $e=0.99$ and $\iota=30^{\circ}$. The snapshots are each of $6000$s duration, and are shown when the semi-latus rectum has the value $p=20$M (top left), $p=10$M (top right), $p=5$M (bottom left) and $p=2.75$M (bottom right).}
\label{snapshots}
\end{figure}

As mentioned earlier, inspirals generated using this hybrid scheme can be used to investigate the qualitative properties of EMRIs, and also to construct approximate, ``kludge'', EMRI waveforms for use in LISA data analysis. By integrating the Kerr geodesic equations along the trajectory through phase space constructed using the method described here, the path followed by a particle can be computed in Boyer-Lindquist coordinates. Identifying these coordinates with spherical polar coordinates and constructing the corresponding flat space quadrupole moment tensor allows the generation of an approximate gravitational waveform. Such waveforms are being used to scope out LISA data analysis algorithms \cite{gair04}. In Figure~\ref{snapshots} we show short snippets of the kludge gravitational waveform from one of the inspirals illustrated in Figures~\ref{peInsp} and \ref{piInsp}, with $a=0.9$M and initial parameters $p=20$M, $e=0.99$ and $\iota=30^{\circ}$. The snippets are of $6000$s duration (assuming masses of $M=10^6$\msun \, and $\mu=1$\msun) and are shown at four points along the inspiral trajectory. We include this figure for illustration purposes only. More details on the construction of kludge waveforms can be found in \cite{kludge_paper}.

%%%%%%%%%%%%%%%%%%%%%%%%%%

\acknowledgments 

We thank Stanislav Babak and Scott Hughes for useful discussions and Scott Hughes for supplying the Teukolsky data used in computing the fits for the radiation from circular orbits. We also thank Bernard Whiting for useful suggestions regarding the evolution of the Carter constant. The work of JRG was supported in part by NASA grants NAG5-12834 and NAG5-10707 and by St.Catharine's College, Cambridge. KG acknowledges support from PPARC Grant PPA/G/S/2002/00038.

%%%%%%%%%%%%%%%%%%%%%%%%%%%%%%%%%%%%%%%%%%%%%%%%%%%%%%%%%%%%%%%%%%%%%%%%%%%%%%%%%%%%%%%%%%%%%%%%%%%%%%%%%%%%%%

\end{document}